\documentclass[10pt,conference]{IEEEtran}
\usepackage{epsfig,setspace,amsmath,epsf,amssymb,bm,cite,graphicx, epstopdf,algorithm,algpseudocode,float,color,mathtools,authblk,tikz,physics,amsthm,enumitem}

\usetikzlibrary{positioning}
\usetikzlibrary{decorations.text}
\usetikzlibrary{decorations.pathmorphing}
\usepackage[short,c2]{optidef}

\usepackage{empheq}
\usepackage{cases}

\usepackage[table,xcdraw]{xcolor}
\usepackage{booktabs}
\usepackage{subfigure}
\usepackage{bbm}
\newcommand{\bracenom}{\genfrac{\lbrace}{\rbrace}{0pt}{}}

\newtheorem{theorem}{Theorem}

\IEEEoverridecommandlockouts
\allowdisplaybreaks
\begin{document}
\title{Age of Information with \\ Age-Dependent Server Selection}

%\title{Minimization of Functions of Age of Incorrect Information in Remote Estimation with DR-AMC}

\author[1]{Nail Akar}
\author[2]{Ismail Cosandal}
\author[2]{Sennur Ulukus}

\affil[1]{\normalsize Bilkent University, Ankara, T\"{u}rkiye}
\affil[2]{\normalsize University of Maryland, College Park, MD, USA}
\maketitle
\begin{abstract}
In this paper, we consider a single-source multi-server generate-at-will discrete-time non-preemptive status update system where update packets are transmitted using {\em only one} of the available servers, according to a server selection policy. 
In particular, when a transmission is complete, the update system makes a threshold-based decision on whether to wait or transmit, and if latter, which server to use for transmissions, on the basis of the instantaneous value of the age of information (AoI) process. In our setting, servers have general heterogeneous discrete phase-type (DPH) distributed service times, and also heterogeneous transmission costs.
The goal is to find an age-dependent multi-threshold policy that minimizes the AoI cost with a constraint on transmission costs, the former cost defined in terms of the time average of an arbitrary function of AoI.  
For this purpose, we propose a novel tool called \emph{multi-regime absorbing Markov chain} (MR-AMC) 
in discrete time. Using the MR-AMC framework, we exactly obtain the distribution of AoI, and subsequently the costs associated with AoI and transmissions.
With the exact analysis in hand, optimum thresholds can be obtained in the case of a few servers, by exhaustive search. We validate the proposed analytical model, and also demonstrate the benefits of age-dependent server selection, with numerical examples. 
\end{abstract}
\section{Introduction}
In status update systems, information update packets carrying sample values of an information source process are transmitted by sources towards remote monitors, using communication links or networks which introduce random delays  
with the goal of keeping the monitor's view of the source as fresh as possible \cite{elmagid_commag19}. 
Therefore, a need is evident for quantifying information freshness in a way different than conventional network performance metrics including delay or loss. 
For this purpose, age of information (AoI) process, or age process, was first introduced in \cite{kaul_etal_infocom12} which is a continuous-time random process keeping track of the elapsed time since the generation time of the last received status update, from the remote monitor's perspective. 
The continuous-time AoI process is composed of AoI cycles during which the process increases at a unit rate within a cycle which ends with the reception of a packet upon which the AoI process is subject to a downward jump. 
We refer the reader to \cite{yates_survey, kosta_etal_survey,abbas_survey23} for surveys on AoI analysis and optimization. In the majority of existing work, time-averaged AoI is sought as the information freshness metric,  whereas more general time averages of arbitrary functions of AoI are also considered \cite{tripathi_modiano_TNET}. 
On the other hand, the AoI process is also studied in discrete time for which the discrete-time AoI process increases by one at every time slot unless a new packet is received, whereas the process drops to the system time of the received packet upon its reception
\cite{kosta_etal_jsac21,akar_dogan_iot21,zhang_etal_tcom25}. 
In this paper, we focus on the AoI process in discrete time, and the time average of an arbitrary function of AoI is used to represent the AoI cost.

Two general frameworks are considered in the AoI literature depending on how status update packets are generated. 
In the random arrival (RA) framework, sampling is done by the sources according to a random process 
without a reference to the transmitter/server status. In RA models, when incoming updates find an ongoing transmission, they can be lost, or buffered for transmission at a later time \cite{yates_kaul_TIT19}, or allowed to preempt the ongoing transmission \cite{dogan_akar_tcom21}.
On the other hand, in the generate-at-will (GAW) framework, see for example \cite{sun_etal_TIT17,akar_ulukus_tcom25}, the transmitter is in charge of deciding when to sample the information source on the basis of the server status, and additionally the instantaneous AoI, for 
transmitting its information packet, at its will \cite{yates_survey}. Although preemption is possible in GAW systems \cite{banerjee_ulukus_isit24}, queuing delays and losses can entirely be avoided.
This paper's focus is on the GAW framework.

There are two types of multi-server systems studied from the AoI perspective, depending on how the servers are collectively utilized for status updates. In the first type which is more common, it is possible to simultaneously use the servers for transmissions. 
The authors of \cite{akar_ulukus_tcom25},\cite{kam_etal_TIT16},\cite{yates_isit18} study AoI for two servers, infinitely many servers, and finite number of servers, respectively, all allowing simultaneous transmissions in various GAW and RA settings. 
In these settings, simultaneous use of multiple servers is shown to enhance the freshness performance. 
However, a drawback of simultaneous transmissions is the emergence of out-of-order packets at the receiver. 
% IZW particular, when a packet's transmission is over, some of the packets in transmission with earlier timestamps than the received packet, become obsolete in terms of AoI. These obsolete packets may be discarded at the monitor while wasting server resources, or they can immediately be preempted by the source to reserve resources. 
% A drawback of this approach is its energy inefficiency stemming from obsolete packets and the need for relatively larger number of transmissions.
In the second type, simultaneous transmissions over multiple servers is not allowed, see for example \cite{purdue_paper}. This paper's focus in on the second type.

For deriving the freshness metrics derived from the AoI process, several approaches exist.
The graph-based approach \cite{kaul_etal_infocom12} enables the calculation of the average AoI using graphical techniques in relatively simpler systems.
On the other hand, the stochastic hybrid systems (SHS) approach is proposed in \cite{yates2019} for obtaining  the average AoI in a systematical way for a single-buffer server receiving packets randomly arriving from multiple sources. The moment generating function (MGF) and also the higher order moments of AoI, have also been studied by the SHS approach in various settings \cite{yates2020age}. 
An alternative technique for AoI modeling is the absorbing Markov chain (AMC) approach proposed in \cite{akar_gamgam_comlet23} to obtain the distribution of AoI in matrix-exponential form for a continuous-time multi-source status update system for a GAW system with heterogeneous service times, and an RA system using a single-packet buffer. 
The AMC method has recently been applied to a single-source dual-server status update system in discrete time \cite{yifan_etal_unpublished25}. 
However, existing approaches for AoI including SHS and AMC have mostly focused on the study of age-agnostic control policies which do not make use of the instantaneous AoI in making decisions regarding transmissions. In this paper, we propose a method to extend the AMC approach originally developed for age-agnostic settings, to a single-source GAW multi-server system employing an age-dependent server selection policy. 
% One of the few analytical models for age-dependent policies can be found in \cite{maatouk_etal_jsac23} which allows the system’s transition dynamics to be a polynomial function of AoI. 
% In this paper, we develop an AMC-based method for obtaining the distribution of AoI of certain closed-loop (or age-dependent) threshold policies for a dual-channel system for which the instantaneous AoI is utilized in making status update decisions.
% on whether to wait or not and if latter, which channel to transmit. Moreover, finding the distribution of AoI makes it possible to obtain the expected value of any function of AoI, which differentiates the current paper from the existing literature.
\begin{figure}
 \centering
\begin{tikzpicture}[scale=0.30,stateW/.style={circle, draw=red!60, fill=red!5, very thick, minimum size=6mm},state1/.style={circle, draw=blue!60, fill=blue!5, very thick, minimum size=6mm},state2/.style={circle, draw=gray!60, fill=gray!5, very thick, minimum size=6mm}]
	%\draw[gray] (0,2) node {x} (0.5,1) node {y} (1,0.75) node {z} (1.5,0.9) node {v} (2,1) node {w} (2.5,0.8) node{k};
	\pgfplothandlercurveto
	\pgfplotstreamstart
	% \pgfplotstreampoint{\pgfpoint{0cm}{6cm}}
	 \pgfplotstreampoint{\pgfpoint{0.2cm}{4.9cm}}
    \pgfplotstreampoint{\pgfpoint{0.5cm}{5.7cm}}
	\pgfplotstreampoint{\pgfpoint{1cm}{6.9cm}}
	\pgfplotstreampoint{\pgfpoint{1.5cm}{6.2cm}}
	\pgfplotstreampoint{\pgfpoint{2cm}{6.3cm}}
	\pgfplotstreampoint{\pgfpoint{2.5cm}{5.5cm}}
	\pgfplotstreampoint{\pgfpoint{3cm}{5.7cm}}
	\pgfplotstreampoint{\pgfpoint{3.5cm}{5.4cm}}
	\pgfplotstreampoint{\pgfpoint{4cm}{5.2cm}}
	\pgfplotstreampoint{\pgfpoint{4.5cm}{5.7cm}}
    \pgfplotstreampoint{\pgfpoint{4.8cm}{5.9cm}}
    \pgfplotstreampoint{\pgfpoint{4.9cm}{5.91cm}}
	\pgfplotstreamend
	\pgfusepath{stroke}
    \draw[rounded corners,thick,black] (15.6,4.7) rectangle (22.6,7.3) {};
	\draw[rounded corners,thick,black] (0,4.5) rectangle (5,7.5) {};
    \filldraw (2.4,8.25) node[anchor=center] {\scriptsize{source process}};
    \filldraw (19.1,6) node[anchor=center] {\scriptsize{remote monitor}};
    \node[state2]  at (10,10.5)   {\scriptsize{Wait}};
    \draw[thick,->,gray] (5.5,7) -- (8.8,9.8) ;
     \node[state1]  at (10,6.5)   {\scriptsize{Server $1$}};
      \draw[thick,->,blue] (5.5,6) -- (8,6.5) ;
       \draw[thick,->,blue] (12.2,6.4) -- (15.5,6.5) ;
   %    \draw[ultra thick,->,blue] (11.5,7) -- (17,6) ;
    \node[]  at (10,3.5)   {\large{$\vdots$}};
     \node[]  at (14,5.5)   {\large{$\vdots$}};
     \node[]  at (7,5.4)   {\large{$\vdots$}};
    \node[stateW]  at (10,0)   {\scriptsize{Server $J$}};
     \draw[thick,->,red] (5.5,5.5) -- (8.5,1.4) ;
      \draw[thick,->,red] (12.1,1) -- (15.4,5.4) ;
   %   \draw[ultra thick,->,blue] (11.5,4) -- (17,5.5) ;
  % \filldraw[blue] (10,2.3) circle (0.075) node[anchor=center] {};
  %  \filldraw[blue] (10,1.8) circle (0.075) node[anchor=center] {};
  %   \filldraw[blue] (10,1.3) circle (0.075) node[anchor=center] {};
  %  \node[state1]  at (10,-0.5)   {\scriptsize{Server $J$}};
   % \draw[ultra thick,->,blue] (5,4.5) -- (8.5,0) ;
    % \draw[ultra thick,->,blue] (11,0.5) -- (17,5) ;
\draw[ultra thick,->] (5.5,7.5) arc (45:-30:3)  node[anchor=south west] {};
\end{tikzpicture}
\caption{Multi-server status update system that makes a decision on whether to wait, or transmit using one of the $J$ servers, when there is no ongoing transmission}
\label{fig:systemmodel}
\end{figure}
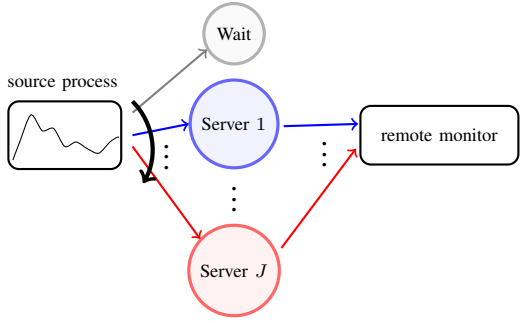
For a single-source single-server GAW status update system, the zero-wait (ZW) policy refers to  one where the source transmits a status update packet immediately after the previous packet service time is complete \cite{sun_etal_TIT17}.
% When the service time of a packet turns out to be relatively short, ZW policy triggers the immediate transmission of a new packet generating two back-to-back packets with very close timestamps, which is not in favor of AoI.
On the other hand, non-zero-wait (NZW) policies are first proposed in \cite{sun_etal_TIT17} for which the source waits for some time before transmitting, where the wait time should depend on the instantaneous value of the AoI. In particular, the authors of \cite{sun_etal_TIT17} propose optimal threshold-based NZW algorithms in a single-server setting. 
% On the other hand, \cite{moltafet_etal_ton23} studies a discrete-time status update system under two-way delay by deriving the average AoI expressions for several ZW and NZW policies. 
In this paper, we study a single-source multi-server NZW status update system in discrete time.  Once the service time of a packet is over, the transmitter makes a threshold-based decision on whether to wait or transmit, as a function of the instantaneous AoI as in \cite{sun_etal_TIT17}. A further threshold-based decision is to be made on which of the servers to use for transmission, on the basis of the instantaneous AoI, leading to a multi-threshold transmission policy. 
This system is illustrated in Fig.~\ref{fig:systemmodel}.

In the current work, servers are associated with heterogeneous discrete phase-type (DPH) distributed service times with unbounded or bounded support \cite{telek_book}, and also heterogeneous transmission costs. Heterogeneity in service times and transmission costs pave the way for the development of age-dependent transmission policies. For example, when instantaneous age is small, a slow server with lower transmission cost can be preferred for status updates.  However, for the contrary scenario of large ages, one may resort to a faster server to bring down the AoI process as quickly as possible, despite its higher transmission costs. 
Our goal is to obtain the thresholds which minimize the AoI cost 
under a constraint on the transmission cost, where the AoI cost is expressed as the time average of an arbitrary function of AoI.  On the other hand, the transmission cost is taken as a weighted sum of server use frequencies.
A novel mathematical tool called \emph{multi-regime absorbing Markov chain} (MR-AMC) in discrete-time is introduced in this work as an extension of the AMC method of \cite{akar_gamgam_comlet23},  
to exactly obtain the distribution of AoI, given the thresholds of the underlying age-dependent policy. The MR-AMC tool is recently used in \cite{cosandal_etal_tit25} in continuous time to derive optimum transmission policies for minimizing the average age of incorrect information (AoII) in a push-based status update system with its information source modeled as as continuous-time Markov chain (CTMC).  
The distribution of absorption time in an MR-AMC in continuous-time, is a sub-case of the inhomogeneous phase-type (PH-type) distribution recently proposed in the field of applied probability (see for example \cite{albrecher2019inhomogeneous},\cite{albrecher2022fitting}) with finitely many regimes. However, the use of inhomogeneous MR-AMCs in discrete time with finitely many thresholds and multiple absorbing states, is novel to this paper, to the best of our knowledge.   
%In the current work, we extend the work of \cite{cosandal_etal_tit25} to discrete time by introducing the MR-AMC in discrete time and then using it to model age-dependent server selection.  
Moreover, finding the distribution of AoI makes it possible to obtain the expected value of any  function of AoI, which differentiates the current paper from the existing literature.
In the case of few servers, one can use brute-force search to find the optimum thresholds. 

Our contributions in the current paper are summarized as follows:
\begin{itemize}
\item We introduce a new stochastic analysis tool, namely a multi-regime absorbing Markov chain, in discrete-time, which is shown to model age-dependent policies exactly for GAW status update systems.
\item In particular, given a set of thresholds characterizing the age-dependent server selection policy, we derive the distribution of AoI using the MR-AMC formulation. Moreover, this distribution being in matrix geometric form for large ages, allows us to find the age violation probabilities, or the time average of a polynomial function of AoI in closed-form. Alternatively, time averages of more general functions of AoI can be obtained numerically.
\item We also derive the transmission costs as a function of the thresholds, using the MR-AMC model.
\end{itemize}

The paper is organized as follows. In Section~\ref{sec:related}, we present the related work. Section~\ref{sec:prel} presents the preliminaries on notation and discrete-time absorbing Markov chains. 
In Section~\ref{sec:mramc}, we introduce MR-AMCs and their properties key to the development of this paper. 
In Section~\ref{sec:systemmodel}, the system model is presented.
Section~\ref{sec:analysis} presents the analytical method for deriving the distribution of AoI, and subsequently obtaining the AoI and transmission costs. Validation of the analytical models and comparative evaluation of various policies is presented in Section~\ref{sec:numerical}. Conclusions, open problems and future research directions are presented in Section~\ref{sec:conclusions}.
\section{Related Work}
\label{sec:related}
The problem of server selection has been extensively studied in various settings \cite{hawking1999methods},\cite{fei_etal_infococom98}. However, there are only a few works focusing on age-minimizing server selection, which are briefed in this section.
The authors of \cite{pan_etal_tmc22} have studied age-optimal transmission
scheduling for dual-server systems with non-renewal service times. 
In particular, transmissions have two options:
an unreliable but fast (e.g., mmWave) channel, or a reliable but slow (e.g., sub-6 GHz) channel.
Age-optimal policies in this setting have been proven to be of threshold-type on the age,
and low complexity algorithms have been developed for finding the optimal scheduling policy. 
Average AoI minimization by server selection is studied in \cite{purdue_paper} for a multi-server system for which the authors 
show that both the optimal waiting time and the optimal server selection policies admit a water-filling structure, which can be computed by a 
fixed-point-based numerical method. However, their method is limited to bounded support service times, and only the average AoI is considered as the AoI cost. 
The authors of \cite{atasayar_etal_mobihoc25} attempt to minimize the average AoI in a scenario where the sender chooses to forward a status update over one of the available routes which have distinct un-bounded support continuous delay statistics, using  a semi-Markov decision process formulation. 
Our work is different from \cite{atasayar_etal_mobihoc25} since we seek the minimization of a more general AoI cost which requires the derivation of the distribution of AoI. 
Moreover, the MR-AMC technique key to the proposed approach is not limited to age-dependent server selection only, and it can also be applied to AoI modeling for other age-dependent policies.
We note that the dual-server sub-problem of the current work is investigated in our earlier work \cite{akar_Asilomar}. 
\section{Preliminaries}
\label{sec:prel}
\subsection{Notation}
Uppercase bold letters are used to denote real-valued matrices whereas uppercase letters denote random variables.
Lowercase bold (plain) letters or symbols, are used to denote real-valued vectors (scalars).
The $(i,j)^{\text{th}}$ element of a matrix $\bm{A}$ and an indexed matrix $\bm{A}_n$ is denoted by $A_{i,j}$ and $A_{n,i,j}$, respectively. 
Similarly, $a_i$ and $a_{n,i}$ stand for the $i^{\text{th}}$ element of a vector $\bm{a}$ and an indexed vector $\bm{a}_n$, respectively.
%The $i^{\text{th}}$ row and $j^{\text{th}}$ column of a matrix $\bm{A}$ is denoted by $\bm{A}(i,:)$ and $\bm{A}(:,j)$, respectively. 
%
The notations $\bm{0}_{m \times n} $, ${\bm I_m}$, and ${\bm 1_m}$ denote a matrix of zeros of size $m \times n$, an identity matrix of size $m$, and a column vector of
ones of size $m$, respectively. When used without a subscript, size information is inferred from the context. 
% The function $u_k$ stands for the discrete-time unit step function, i.e., $u_k=1,k=0,1,\ldots$ and is zero otherwise. The function $\delta_k$ stands for the discrete-time unit impulse function, i.e., $\delta_k=1$ for $k=0$ and is zero, otherwise.
\subsection{Discrete Phase-type Distribution}
\label{subsection:dph}
A discrete phase-type (DPH) distributed random variable is defined as the time until 
absorption, denoted by $T$, in a finite-state discrete-time
Markov chain (DTMC) $X_k \in \{ 1,\ldots, M,M+1 \}, \ k=0,1,\ldots,$ with the first $M$ states being transient states, and the last state $M+1$ designated as the absorbing state \cite{alfa_book,nielsen_book,telek_book}. The DTMC $X_k$ has the initial probability vector of size $1 \times M$ denoted by $\bm{\beta}$, 
\begin{align}
    \bm{\beta} & = \begin{pmatrix}
     \beta_1 &   \cdots & \beta_M 
    \end{pmatrix}, \ \beta_i = \Pr (X_0 = i), i=1,\ldots,M, 
\end{align}
and probability transition matrix $\bm Q$ of the form,
\begin{align}
\bm{Q} & = \left( \begin{array}{c|c} \bm{A} & \bm{a} \\ \hline  \bm{0} & 1 \end{array} \right), \label{absorbing}
\end{align}
for a sub-stochastic matrix $\bm{A}$ called the transient matrix, and a column vector $\bm{a}=\bm{1}-\bm{A}\bm{1}$, called the absorption vector.  In this case, we say $T \sim \text{DPH}(\bm{\beta},\bm{A})$ with order $M$. We write the transient probability vector of the AMC $X_k$ of size $1 \times M$ at time $k$ as follows,
\begin{align}
   \bm{x}_k & = \begin{pmatrix}
      x_{k,1} & x_{k,2} & \cdots & x_{k,M}  
   \end{pmatrix}, \;
   x_{k,m}   = \Pr (X_k=m), \label{xki}
\end{align} 
which then leads to the following closed-form expression for $\bm{x}_k$,
\begin{align} \bm{x}_k & = \bm{\beta} \bm{A}^k, k \geq 0. \end{align} 
The absorption time $T$ has  cumulative distribution function (cdf) $F_T(n)= \Pr (T \leq n)$, and probability mass function (pmf) $p_T(n)=\Pr (T =n)$, which can respectively be written for $n=1,2,\ldots$, 
\begin{align}
F_T(n) & = 1 - \bm{x}_n \bm{1} = 1 - \bm{\beta} \bm{A}^n \bm{1}, \label{eq:cdf} \\
p_T(n) & = F_T(n) - F_T(n-1) = \bm{\beta} \bm{A}^{n-1} \bm{a}. \label{eq:pdf} 
\end{align}
 Moreover, the $i^{\text{th}}$ factorial moment of $T$, denoted by $\nu_{i}, i=1,2,\ldots$, can be written in closed form through the following expression \cite{telek_book},
\begin{align}
   \nu_{i} & = \mathbb{E}[\, T  (T-1) \ \cdots \ (T-i+1) \,],  \\ 
   & = \bm{\beta} \left( \sum_{n=1}^{\infty} n(n-1)\cdots (n-i+1) {\bm A}^{n-1} \right) \bm{a}, \\
   & = \bm{\beta}  \ \left( i! \,  (\bm{I}-\bm{A})^{-i-1} \bm{A}^{i-1} \right) \ \bm{a}.
   \label{factorialMoments}
\end{align}
Additionally, the $i^{\text{th}}$ ordinary moment of $T$, denoted by $\mu_i=\mathbb{E}[T^i], i=1,2,\ldots$, can be obtained from the factorial moments \cite{bagui2024stirling} by,
    \begin{align}
        \mu_i & = \bm{\beta} \left( \sum_{n=1}^{\infty} n^i {\bm A}^{n-1} \right) \bm{a} = \sum_{j=0}^i \bracenom{i}{j} \nu_j, \label{eq:relation} \\
         & = \bm{\beta}  \ \left( \sum_{j=0}^i j! \ \bracenom{i}{j}  (\bm{I}-\bm{A})^{-j-1} \bm{A}^{j-1} \right) \ \bm{a}, \label{eq:closedform}
    \end{align}
    where $\bracenom{i}{j}$ stands for the Stirling number of the second kind. As a result, when the distribution of AoI is given in a matrix-geometric form similar to \eqref{eq:pdf}, then the time average of a polynomial function of AoI can be obtained in closed-form using the procedure given in the identities \eqref{eq:relation} and  \eqref{eq:closedform}.
% The $i^{\text{th}}$ ordinary moment of $T$, denoted by $\mu_{i} =  \mathbb{E}[T ^{i}]$, can be obtained from the factorial moments $\gamma_j, \ 1 \leq j \leq i$ given in closed form in \eqref{factorialMoments}, but also more efficiently from the following recurrence derived in \cite{dayar_akar_siam05}
% with $\bm{m_0}=\bm{1}_{M \times 1}$,
% \begin{align}
%    (\bm{I}-\bm{A}) \; \bm{m}_{i+1} &  = \sum_{j=0}^{i} (-1)^{i-j} \binom{i+1}{j} \bm{m}_i, \ i \geq 1, \\
%    \mu_i & = \bm{\beta} \, \bm{m}_i.
%    \label{ordinaryMoments}
% \end{align}
DPH distributions are very general and they include the deterministic distribution, uniform distribution, and un-bounded support distributions such as the geometric distribution and 
mixed-geometric distribution, as its sub-cases \cite{alfa_book}.
For example, if $T$ is geometrically distributed with parameter $p$, then $T \sim \text{Geo}(p) \sim \text{DPH}(1,1-p)$ with order one. 
When $T$ is composed of a mixture of two geometrically distributed variables with parameters $p_1$ and $p_2$, mixing weights $w_1$ and $w_2$, we say 
$T \sim \text{MG}(p_1,p_2,w_1,w_2) \sim \text{DPH}\left( \bigl( \begin{smallmatrix}
  w_1& w_2 
\end{smallmatrix} \bigr), \bigl( \begin{smallmatrix}
  1-p_1 & 0\\
  0 & 1-p_2
\end{smallmatrix} \bigr) \right)$.
As the final example, consider a discrete positive random variable $T$ with bounded support pmf, i.e., $p_T(m) \neq 0$ when $m=M$, is zero when $m>M$ and $m=0$, is of DPH-type of order $M$, i.e., $T \sim \text{DPH}(\bm{\beta},\bm{A})$, where $\bm{\beta}$ is a row vector of zeros except for the first element which is one, $\bm{A}$ is a 
square matrix of zeros except for the $(m,m+1)^{\text{th}}$ position which is written as \cite{alfa_book},
\begin{align}
A_{m,m+1} & = 1-\frac{p_T(m)}{\sum_{\ell=m}^M p_T(\ell)}, \ m=1,\ldots,M-1. 
\end{align}
\section{Discrete-time Multi-regime Absorbing Markov Chains}
\label{sec:mramc}
In this section, we describe the MR-AMC, and also the related multi-regime DPH (MR-DPH) distribution. MR-AMC and MR-DPH are generalizations of the DPH distribution and its associated DTMC process 
$X_k, \, k=0,1,\ldots$, with
\begin{align*}
    X_k \in \{ 1,2,\ldots,M,M+1,\ldots,M+J \},
\end{align*} 
with the first $M \geq 1$ states being transient, the last $J \geq 1$ states being absorbing (as opposed to one absorbing state), initial probability vector $\bm{\beta}$ of size $1 \times M$, and the transient and absorption matrices of the DTMC process depending on the regime associated with the elapsed time since the AMC starts evolution, in a piece-wise constant manner. In particular, regime-$i$ is defined as the elapsed time interval $[\tau_{i-1},\tau_{i})$ where $\tau_i,i=1,\ldots,I-1, \ \tau_i < \tau_{i+1},$ are the finite thresholds with $\tau_0 = 0$ and $\tau_{I}=\infty$. During regime-$i$, the transitions among the transient states is governed by the sub-stochastic transient matrix $\bm{A}_i$, and the absorption matrix in regime-$i$ is denoted by $\bm{B}_i$ where 
\begin{align}
    \bm{B}_i\bm{1} & = \bm{1} - \bm{A}_i \bm{1}, \label{bmB1}
\end{align}
where $\bm{A}_i$ is square and of size $M$, and $\bm{B}_i$ is $M \times J$.
Subsequently, in regime-$i$, the probability transition matrix of the DTMC $X_k$ is denoted by $\bm{Q}_i$, which can be written as,
\begin{align}
\bm{Q}_i & = \left( \begin{array}{c|c} \bm{A}_i & \bm{B}_i \\ \hline \bm{0} & \bm{I} \end{array} \right). \label{absorbing_regimei}
\end{align}
In particular, the fundamental matrix $(\bm{I}-\bm{A}_I)^{-1}$ exists which ensures that the chain is guaranteed to absorb into an absorbing state.
% Moreover, the $j$th column of $\bm{B}_i$, for $j=1,\ldots,J$, represents the absorption vector into absorbing state-$j$ at regime-$i$, and is denoted by  
% $\bm{B}_{i,j}$.
We define the absorption vector in regime-$i$, denoted by $\bm{\sigma}_i$, 
\begin{align}
   \bm{\sigma}_i & = \begin{pmatrix}
      \sigma_{i,1} & \sigma_{i,2} & \cdots & \sigma_{i,J}  
   \end{pmatrix}, 
   \end{align}
where $\sigma_{i,j}$ is the probability of absorption into absorbing state-$j$ stemming from a transition taking place when the elapsed time is in regime-$i$. The following theorem states our main result on MR-AMCs, which is needed for the development of this paper, and its proof is given in Appendix~\ref{appendix:A}.  
\begin{theorem} 
\label{theorem:thm1} 
Consider 
the MR-AMC $X_k$ with $M$ transient states, $J$ absorbing states, $I$ regimes defined through the $I-1$ finite thresholds 
$\{ \tau_i \}_{i=1}^{I-1}$, and the probability transition matrix in regime-$i$, $\bm{Q}_i$, written as in \eqref{absorbing_regimei}.
Then, the cdf of the absorption time $T$ of the MR-AMC can be written as, 
\begin{align}
    F_T(n) & = 1 - \bm{\beta}_i \bm{A}_i^{n-\tau_{i-1}} \bm{1}, \ \tau_{i-1} \leq n < \tau_{i}, \label{timetoabsorb}
\end{align} 
where
 \begin{align}
    \bm{\beta}_1 & = \bm{\beta}, \ \bm{\beta}_i = \bm{\beta}_{i-1} \bm{A}_{i-1}^{\delta_i},  \ \delta_i = {\tau_i - \tau_{i-1}}. \label{betaevolution}
\end{align}    
Moreover, the absorption vector in regime-$i$ is written in closed-form as,
\begin{align}
 \bm{\sigma}_{i} & = \bm{\beta}_i 
% (\bm{I}-  \bm{A}_i^{\delta_i}) (\bm{I} - \bm{A}_i)^{-1} \sum_{l=0}^{\delta_i - 1} \bm{A}_i ^l
\left( \sum_{l=0}^{\delta_i - 1} \bm{A}_i ^l \right) 
\bm{B}_{i}, \ 1 \leq i \leq  I. \label{absorptionall} 
\end{align} 
In particular, the absorption vector in regime-$I$ can be written in closed-form,
\begin{align}
  \bm{\sigma}_{I} & = \bm{\beta}_I (\bm{I} - \bm{A}_I)^{-1} \bm{B}_{I}. \label{absorptionfinalregime}
\end{align}
\end{theorem}
 The MR-AMC $X_k$ and its absorption time $T$ of the MR-AMC are characterized with the 4-tuple, 
\begin{align}
   \left( \bm{\beta},\{ \tau_i \}_{i=1}^{I-1},\{ \bm{A}_i \}_{i=1}^{I}, \{ \bm{B}_i \}_{i=1}^{I} \right).  \label{4tuple}
\end{align}
In this case, we say that $T$ has an MR-DPH distribution characterized with the 4-tuple \eqref{4tuple}. Also note that the original DPH distribution is a sub-case of MR-DPH for which there is a single regime, i.e., $I=1$, and all $J$ absorbing states are merged into one absorbing state.

\section{System Model}
\label{sec:systemmodel}
We consider a non-preemptive status update system consisting of a single information source and a remote monitor.
The information source samples a corresponding random process at times according to
the GAW principle, and transmits the sampled values towards the monitor using information packets, while using one of the $J \geq 1$ available servers for transmissions. 
The service time of an information packet for the $j^{\text{th}}$ server, denoted by $S_j, j=1,\ldots,J$, is assumed to have a DPH distribution. In particular, $S_j \sim \text{DPH}(\bm{\alpha}_j,\bm{D}_j)$ with order $M_j$, and its absorption vector is denoted by $\bm{d}_j = \bm{1} - \bm{D}_j\bm{1}$. A transmission of a packet on server-$j$ is assumed to have a transmission cost $c_j$, which depends on the server type. 

The ordering of events at time slot $k$ is now described. For this purpose, we first define $P_{k-1}$ as the information packet which was under transmission during previous slot $k-1$. 
\begin{enumerate}[leftmargin=0.6cm]
    \item In the first step, the source checks whether the ongoing transmission of packet $P_{k-1}$ is complete or not.  \item In the second step, the source updates the discrete-time AoI process $\Delta_k$ according to,
\begin{align}
    \Delta_k &=  
    \begin{cases}
   k - g_{P_{k-1}}, & \text{if $P_{k-1}$ just received,}  \\
      \Delta_{k-1} +1,    & \text{otherwise,}       
    \end{cases}
\end{align}
where $g_{P_{k-1}}$ is the generation time of the packet $P_{k-1}$ which is just received. 
Note that the age process is incremented if there was no ongoing transmission or the ongoing transmission of $P_{k-1}$ does not get to complete. 
\item In the third step, we apply an age-dependent packet transmission policy called $\mathcal{P}$ characterized in terms of the $J$ thresholds 
$\{ \tau_i \}_{i=1}^{J},\tau_0=0,\tau_{J+1}=\infty$,
used by the source. In particular, if there is an ongoing transmission, the source stays idle.
% Note that preemption of packets in transmission is not assumed in this work.
If there is no ongoing transmission, and if $\Delta_k < \tau_1$, then the source stays idle,
else if $\tau_1 \leq \Delta_k \leq \tau_2$, then server-$1$ is used, else if 
 $\tau_j < \Delta_k \leq  \tau_{j+1}$, then server-$j$ is used, for sampling and transmitting an update packet. Note that $\tau_1$ may be equal to $\tau_2$ but otherwise $\tau_j > \tau_{j-1}$.
\end{enumerate}
In this setting, we assume that the source has full knowledge of the instantaneous value of the AoI process $\Delta_k$ at all times due to immediate acknowledgment of the completed packet at the end the first step. The notation $\Delta$  denotes the steady-state random variable for the random process $\Delta_k$ with 
pmf
denoted by $p_{\Delta}(\cdot)$, i.e., 
\begin{align}
p_{\Delta}(n) & = \lim_{k \rightarrow \infty} \Pr ( \Delta_k = n), \ n=1,2,\ldots.
\end{align}
Given the policy $\mathcal{P} \sim \{ \tau_j \}_{j=1}^J$, the main goal of this paper is to derive $p_\Delta(n)$, which enables us to find the expected value of any function of AoI, which is named as AoI cost.
Assuming ergodicity of the process $\Delta_k$ in the general sense, and an arbitrary function $f(\Delta_k)$ of the AoI process, the following identity ties the AoI cost $C_{A} = \mathbb{E}[f(\Delta)]$ to the pmf of AoI,
\begin{align}
   C_{A}  &= \sum_{n=1}^{\infty} f(n) \, p_{\Delta}(n) = \lim_{K \rightarrow \infty} \frac{1}{K} \sum_{k=1}^K f(\Delta_k). \label{AoICost}
\end{align}
Let $f_j$ denote the frequency of server-$j$ transmissions, i.e.,
\begin{align}
   f_j  &= \lim_{K \rightarrow \infty}  \frac{1}{K} \sum_{k=1}^K a_{k,j}, \label{frequency}
\end{align}
where $a_{k,j}$ is one if server-$j$ is selected for transmission at time slot $k$. 
Then, the time-averaged transmission cost $C_T$ can be written as,
\begin{align}
    C_T &= \sum_{j=1}^{J} c_j f_j.   \label{TransmissionCost}
\end{align}
The second goal of this paper is to obtain $C_T$ given a multi-threshold policy. Once $C_A$ and $C_T$ are obtained for a given policy, one can then use search to find the optimum thresholds that minimize the AoI cost under a transmission cost constraint. 

\begin{figure}[tbh]
    \centering
    \begin{tikzpicture}[scale=0.25]	
   % \draw[<->,black] (9,13) -- (16,13);
    % \filldraw (12.5,13) circle (0.01) node[anchor=south, thick] {cycle-$l$};
    \draw[thick,->] (0,0) -- (24,0) node[anchor=west] {$k$};
    \draw[thick,->] (0,0) -- (0,13) node[anchor=east] {$\Delta_k$};
     \draw[thin,dotted,gray] (1,0) -- (1,12);  
      \draw[thin,dotted,gray] (2,12) -- (2,0)  node[anchor=north] {\scriptsize{2}};  
      \draw[thin,dotted,gray] (3,0) -- (3,12); 
      \draw[thin,dotted,gray] (4,0) -- (4,12) ;
       \draw[thin,dotted,gray] (5,0) -- (5,12) ; 
      \draw[thin,dotted,gray] (6,0) -- (6,12)  ;
      \draw[thin,dotted,gray] (7,0) -- (7,12) ;
      \draw[thin,dotted,gray] (8,0) -- (8,12) ;
      \draw[thin,dotted,gray] (9,12) -- (9,0) node[anchor=north] {\scriptsize{9}};  
      \draw[thin,dotted,gray] (10,0) -- (10,12);  
      \draw[thin,dotted,gray] (11,0) -- (11,12); 
      \draw[thin,dotted,gray] (12,0) -- (12,12) ;
       \draw[thin,dotted,gray] (13,0) -- (13,12) ; 
      \draw[thin,dotted,gray] (14,12) -- (14,0) node[anchor=north] {\scriptsize{14}} ;
      \draw[thin,dotted,gray] (15,0) -- (15,12) ;
      \draw[thin,dotted,gray] (16,12) -- (16,0) node[anchor=north] {\scriptsize{16}}; 
      \draw[thin,dotted,gray] (17,12) -- (17,0) ;
      \draw[thin,dotted,gray] (18,0) -- (18,12) ;
       \draw[thin,dotted,gray] (19,0) -- (19,12) ; 
      \draw[thin,dotted,gray] (20,0) -- (20,12)  ;
      \draw[thin,dotted,gray] (21,0) -- (21,12) ;
      \draw[thin,dotted,gray] (22,0) -- (22,12) ;
       \draw[thin,dotted,gray] (23,0) -- (23,12) ;
     % \draw[thin,dotted,gray] (24,0) -- (24,12) ;
   %  \draw[ultra thick,red] (4.5,4.5) -- (9,9);
%   \draw[red,thin] (9,10) circle (7pt);
   \filldraw[red] (8,9) circle (7pt);
   \filldraw[red] (7,8) circle (7pt);
   \filldraw[red] (6,7) circle (7pt);
   \filldraw[red] (5,6) circle (7pt);
   \filldraw[red] (4,5) circle (7pt);
   \filldraw[red] (3,4) circle (7pt);
   \filldraw[red] (2,3) circle (7pt);
\filldraw[blue] (1,2) circle (7pt);
   \filldraw[blue] (0,1) circle (7pt);

    % \draw[dotted,very thin] (4.5,4.5) -- (9,9);
    % \filldraw[] (4,4) circle (2pt);
    % \filldraw[] (3.5,3.5) circle (2pt) ;
    % \filldraw[] (3,3) circle (2pt); 
   % \draw (1.5,5) node[anchor=east] {$u_l$};
  % \draw[dotted,gray] (1.5,8) -- (23,8);
    \draw[dashed,gray,thick] (0,6) -- (23,6);
    \draw[dashed,gray,thick] (0,3) -- (23,3);
     \draw[dotted,gray,thin] (0,1) -- (23,1);
    \draw[dotted,gray,thin] (0,3) -- (23,3);
     \draw[dotted,gray,thin] (0,4) -- (23,4);
    \draw[dotted,gray,thin] (0,5) -- (23,5);
    \draw[dotted,gray,thin] (0,7) -- (23,7);
     \draw[dotted,gray,thin] (0,8) -- (23,8);
    \draw[dotted,gray,thin] (0,9) -- (23,9);
       \draw[dotted,gray,thin] (0,10) -- (23,10);
     \draw[dotted,gray,thin] (0,11) -- (23,11);
    \draw[dotted,gray,thin] (0,12) -- (23,12);
   %  \draw[dotted,gray] (9,12.5) -- (9,0)  node[anchor=north, thick, black] {$d_{l}$};
    % \draw[ultra thick,red] (9,9) -- (9,5.2);
    % \draw[ultra thick,red] (9.1,5.1) -- (16,12);
    % \draw[dotted,black,very thin] (16,12) -- (4,0)  node[anchor=north, thick, black] {$t_{l}$};

    %  \draw[dotted,gray] (0,11) -- (23,11);
    % \draw (1.5,11) node[anchor=east] {$\Phi_l$};
     \draw (0,6) node[anchor=east] {$\tau_2=6$};
    \draw (0,3) node[anchor=east] {$\tau_1=3$};
   %  \draw[dotted,gray] (16,12.5) -- (16,0)  node[anchor=north, thick, black] {$\quad d_{l+1}$};
   
    % \draw[ultra thick,red] (16,12) -- (16,2.2);
    % \draw[ultra thick,red] (16.1,2.1) -- (20,6);
   %  \draw[dotted,black,very thin] (20,6) -- (14,0)  node[anchor=north, thick, black] {${t_{l+1}} \quad$};
    
    \filldraw[green] (9,7) circle (7pt);
    \filldraw[green] (10,8) circle (7pt);
    \filldraw[green] (11,9) circle (7pt);
    \filldraw[green] (12,10) circle (7pt);
    \filldraw[green] (13,11) circle (7pt);
  %  \draw[green,thin] (14,12) circle (7pt);
    % \filldraw[green] (14,12) circle (7pt);
    %\filldraw[green] (15,13) circle (7pt);
    % \draw[green,thin] (16,14) circle (7pt);

%    \filldraw[red] (14,4) circle (7pt);
\filldraw[red] (14,5) circle (7pt);
\filldraw[red] (15,6) circle (7pt);
% \draw[red,thin] (16,7) circle (7pt);

\filldraw[blue] (16,2) circle (7pt);
   \filldraw[red] (17,3) circle (7pt);
    \filldraw[red] (18,4) circle (7pt);
    \filldraw[red] (19,5) circle (7pt);
   \filldraw[red] (20,6) circle (7pt);

    \filldraw[] (22,8) circle (2pt);
    \filldraw[] (21,7) circle (2pt) ;
    \filldraw[] (21.5,7.5) circle (2pt); 
    
% \filldraw[red] (17,7) circle (7pt);
% \filldraw[red] (18,8) circle (7pt);  
    \end{tikzpicture}
    \caption{Sample path of the AoI process $\Delta_k$ for a two-server system when $\tau_1=3,\tau_2=6$. Blue circles indicate the time epochs when the link is idle. Red and green circles indicate transmission epochs using server-$1$ and server-$2$, respectively. }
    \label{fig:samplepath}
\end{figure}
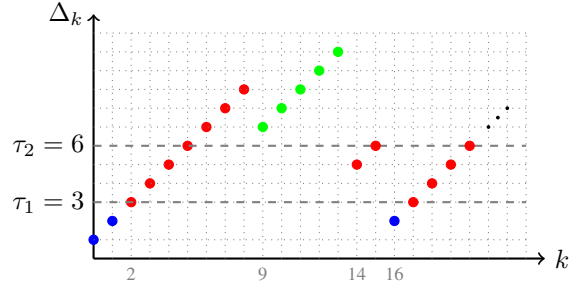

A sample path of the discrete-time AoI process $\Delta_k$ is given in Fig.~\ref{fig:samplepath} for 
a two-server system when $\tau_1=3,\tau_2=6$. The sample path starts from the initial condition $\Delta_0=1$ at which time point there was no ongoing transmission. The source postpones its transmission to $k=2$ when $\Delta_2=3$ at which point a transmission is kicked off at server-$1$ with a service time of 7. Therefore, at time slot $k=9$, the  packet completes and $\Delta_9$ is updated to 7. Since $\Delta_9 > \tau_2$, a transmission is initiated on server-$2$ with a service time of 5.  Consequently, at time slot $k=14$, the service of the packet completes and $\Delta_{14}$ is updated to 5. Since $\tau_1 \leq \Delta_{14} \leq \tau_2$, a transmission is initiated on server-$1$ with a service time of 2
which yields $\Delta_{16}=2 < \tau_1$ so transmission is postponed to $k=17$ at which point a new transmission is kicked off at server-$1$.
% On the other hand, the notation $\Phi$  denotes the steady-state random variable for the random process $\Phi_l$ with pmf 
% $p_{\Phi}$, i.e., $p_{\Phi}(n) = \lim_{l \rightarrow \infty} \Pr ( \Phi_l = n), \ n=1,2,\ldots$

\section{Analytical Model}
\label{sec:analysis}
% In this section, two thresholds $\tau_l$ are used for $l=1,2$, along with $\tau_0 = 0$ and $\tau_3=\infty$ giving rise to $I=3$ regimes where regime-$i$ refers to the interval $[\tau_{i-1},\tau_i)$ in line with the notation used in Section~\ref{sec:systemmodel}.
For AoI analytical modeling of the single-source multi-server update system employing an age-dependent transmission policy, we propose to construct an MR-AMC, namely $X_k, \ k \geq 0,$ which starts operation at $k=0$ with the initiation of transmission of an information packet, called $P_1$. This AMC is allowed to evolve until the service completion of the next packet, called $P_2$, at which time point absorption occurs. 
Recall from Fig.~\ref{fig:samplepath} that the corresponding AoI cycle begins with the reception epoch of $P_1$ and ends just one slot before the reception epoch of $P_2$. Therefore, after $P_1$ is received, the decision on whether we will wait or transmit, and if latter which of the $J$ servers is to be employed, will be made according to the elapsed time of the MR-AMC. When the service time of $P_2$ is over, $X_k$ is to be absorbed into the absorbing state-$j$, if $P_2$ is served by server-$j$. 
The state space of the MR-AMC $X_k$ is given as,
\begin{align*}
    X_k \in \{ (n,j,m) \} \cup \{1,2,\ldots,J\}, 
\end{align*}
where $n=1,2, \ j=1,\ldots,J$, and $m=1,\ldots,M_l$.
The transient state $(n,j,m)$ refers to the situation when packet $P_n$ is served by (or to be served by) server-$j$, and the server process for server-$j$ is in phase $m$, at the beginning of a time slot. Once the absorbing state-$j$ is reached, $X_k$ stays in this state forever.
We define the thresholds of the MR-AMC using the same notation of Section~\ref{sec:mramc} 
% as follows,
% \begin{align}
%      \tau_1 = \gamma_1 +1, \ \tau_j = \gamma_{j}, \ 2 \leq j \leq J, \label{taui}
% \end{align}
% and $\tau_0=0, \ \tau_{J+1}=\infty, \ \delta_i = \tau_i - \tau_{i-1}, \ 1 \leq j \leq J$. 
with regime-$i$ referring to the elapsed time interval $[\tau_{i-1},\tau_i)$ with a total number of $I=J+1$ regimes.
% Depending on whether $P_i$ will be served, or currently served, by server-$l$, we have the transient state $(n,l,m),n=1,2, \ l=1,\ldots,L, \ m=1,\ldots,M_l$ and the absorbing state   
Enumerating all the states of $X_k$ in the following order: 
\begin{align*}
(1,1,1), \ldots, (1,1,M_1), (1,2,1),\ldots,(2,J,M_J), 1, 2, \ldots, J,
\end{align*}
the MR-AMC $X_k$ behaves according to the probability transition matrix $\bm{Q}_i$ in the form \eqref{absorbing_regimei} for regime-$i$, $i=1,\ldots,I=J+1$, with $\bm{A}_i$ being square of size $M$ and $\bm{B}_i$ being of size $M \times J$, 
where 
\begin{align*}
    M & = 2 \sum_{j=1}^J M_j.
\end{align*}
In particular, for regime-$1$, the transient matrix $\bm{A}_1$ is written as,
\begin{align}
 \bm{A}_1 =  \left(  \begin{array}{ccc|ccc}
        \bm{D}_1 & & &   \bm{d}_1 \bm{\alpha}_1 & & \\
     %    &  \bm{D}_1  & & & \bm{d}_2 \bm{\alpha}_1 & & & \\
          &  \ddots & & \vdots & &  \\
        &   &  \bm{D}_J &  \bm{d}_J\bm{\alpha}_1 & &  \\ \hline
        & & &  \bm{I}_{M_1} & &  \\
        & & & & \ddots &  \\
        & & & & &   \bm{I}_{M_L}        
    \end{array} \right), \label{DefA1}
\end{align}
and the absorption matrix $\bm{B}_1$ is a matrix of zeros. Note that empty entries in \eqref{DefA1} 
correspond to matrix blocks of zeros.
On the other hand, the transient matrix for regime-$i$, $i=2,\ldots,I$, can be written as,
\begin{align}
 \bm{A}_i & =   \left( \begin{array}{ccc|ccc|c}
        \bm{D}_1 & & &    & & \bm{d}_1 \bm{\alpha}_{i-1}&  \\
      %  &  \bm{D}_2  & & & & & \bm{d}_2 \bm{\alpha}_{1-1} &  \\
          &  \ddots & & & & \vdots  & \\
        &   &  \bm{D}_J &   & & \bm{d}_J\bm{\alpha}_{i-1} &  \\ \hline
        & & & \bm{D}_1 & & &  \\
         & & &  &  \ddots & &  \\
          & & & & &   \bm{D}_{i-1} &  \\ \hline
        & & & & & &  \bm{I}_{M^{(i)}}        
    \end{array} \right), \label{DefAi}
    \end{align}
where $M^{(i)}=\sum_{j=i}^J M_j$, and the absorption matrix $\bm{B}_i$ can be written as,
    \begin{align}
    \bm{B}_i & = \left( \begin{array}{ccc|cc}
    &  & &  & \\ 
    & & & & \\ \hline
    \bm{d}_1 & & & &\\
    & \ddots & & & \\
    & & \bm{d}_{i-1} & & \\ \hline
    &  & & & 
     \end{array} \right), \label{DefBi}
\end{align}
where the northwest (resp. southwest) block of all zeros of $\bm{B}_i$ has
$M$ (resp. $M^{(i)}$ rows).  
We now describe the evolution of the MR-AMC $X_k$.  
The AMC starts operation at time $k=0$, i.e., at regime-$1$. Let us assume that the MR-AMC starts at state $(1,j,\cdot)$, i.e., $P_1$ starts to receive service from server-$j$, $1 \leq j \leq J$, in some service time phase. When in regime-$1$, 
\begin{itemize}
    \item A transition from phase $(1,j,m)$ to $(1,j,m')$ is incurred with probability ${D}_{j,m,m'}$. Note that these transitions are indicated in the matrix $\bm{D}_j$ in the northwest block of $\bm{A}_1$.
\item While at state $(1,j,m)$, the service of $P_1$ can complete which occurs with probability ${d}_{j,m}$ in which case we need to transition to state $(2,1,m')$ with probability ${\alpha}_{1,m'}$ since $P_2$ will eventually be served by server-$1$ when the threshold $\tau_1$ is reached. 
Note that these 
transitions are reflected in the rank-1 matrices $\bm{d}_j \bm{\alpha}_{1}$ appearing in $\bm{A}_1$.
\item 
When at state $(2,1,m)$, service cannot be started until the threshold $\tau_1$ is reached. Therefore, we continue to stay at $(2,1,m)$ with probability one as long as we are in regime-$1$, which justifies the first identity matrix $\bm{I}_{M_1}$.
\item It is not possible to be in state $(2,j,\cdot)$ in regime-$1$ for $j>1$. Therefore, the outgoing transitions from these states are immaterial.
\item Since the service of $P_2$ cannot start in regime-$1$, the absorption matrix $\bm{B}_1$ is composed of all zeros.
\end{itemize} 
When in regime-$i$, $1 < i \leq I$,
\begin{itemize}
    \item If the service of $P_1$ is ongoing, we should be at state $(1,j,\cdot)$ for some phase in which case:  
    \begin{itemize} 
    \item A transition from phase $(1,j,m)$ to $(1,j,m')$ is incurred with probability ${D}_{j,m,m'}$. Note that these transitions are indicated in the matrix $\bm{D}_j$ in the northwest block of $\bm{A}_i$.
    \item While at state $(1,j,m)$, the service of $P_1$ can complete which occurs with probability ${d}_{j,m}$ in which case we need to transition to state $(2,i-1,m')$ with probability ${\alpha}_{i-1,m'}$ since the service of $P_2$ will be kicked off on server-$(i-1)$. Note that these transitions are reflected in the rank-1 matrices $\bm{d}_j \bm{\alpha}_{i-1}$ appearing in $\bm{A}_i$, along with their location.
    \end{itemize} 
    \item If the service of $P_2$ is ongoing, we should be at state $(2,j,\cdot)$ for $j < i$ for some service phase in which case: 
    \begin{itemize}
    \item A transition from phase $(2,j,m)$ to $(2,j,m')$ is incurred with probability ${D}_{j,m,m'}$. Note that these transitions are indicated in the matrix $\bm{D}_j$ in the middle block of $\bm{A}_i$.
    \item While at state $(2,j,m)$ for $j < i$, the service of $P_2$ can complete in which case we need to transition to absorbing state-$j$ which occurs with probability ${d}_{j,m}$. Note that these transitions are reflected in the term $\bm{d}_j$ for $j<i$ in the absorption matrix $\bm{B}_i$.
    \item It is not possible to be in state $(2,j,\cdot)$ in regime-$i$ for $j>i$. Therefore, the outgoing transitions from these states are immaterial. For this purpose, the southeast block of $\bm{A}_i$ is set to the identity matrix.    \end{itemize}
\end{itemize}
% Once we are in regime-$2$, service is received from server-$1$ for $P_2$ at state $(2,1)$ and a transition to the absorbing state $1$ occurs with probability $p_1$. This situation stays the same in Regime-$3$. For the second scenario, service of $P_1$ can either complete in regime-$2$ in which case we transition $(2,1)$ immediately without wait, or can complete in regime-$3$, upon which
% we transition to state $(2,2)$. From state $(2,2)$, a transition to absorbing state 2 occurs with probability $p_2$.
% The situation related to $P_1$ being served by server-$2$ is also similar. 
% For the DPH-distributed service times, it is clear that we have also transitions among the phases of the individual service times and absorption probabilities are replaced with absorption vectors.
% With this description, we have an MR-AMC described in sub-section~\ref{subsection:MRAMC} with $M=2(M_1+M_2)$ transient states, $J=2$ absorbing states, and $I=3$ regimes. At this point, we have completely characterized the MR-AMC  
% by the 4-tuple $(\bm{\beta},\{ \bm{A}_i \}_{i=1}^{3}, \{ \bm{B}_i \}_{i=1}^{3}, \{ \tau_i \}_{i=1}^{2})$. 
Given the transmission policy $\mathcal{P}$, the information packet $P_1$ is served on server-$j$ with probability $\kappa_j$,
whose value is not known yet. 
However, the relationship between $\bm{\kappa}$ and the initial probability vector $\bm{\beta}$ can be written as,
\begin{align}
\bm{\beta} & =  \begin{pmatrix}
\kappa_1 \bm{\alpha}_1  & \cdots & \kappa_J\bm{\alpha}_J& \bm{0}_{1 \times M}
\end{pmatrix}, \\ 
& = 
\underbrace{
\begin{pmatrix}
\kappa_1  & \cdots & \kappa_J
\end{pmatrix}}_{\bm{\kappa}}
\underbrace{
\left( 
\begin{array}{ccc|c}
\bm{\alpha}_1 & & &  \\
% & \bm{\alpha}_2  & & &  \\ 
& \ddots & & \bm{0}_{J \times M} \\
 & & \bm{\alpha}_J&  \\
\end{array}  \right)}_{\bm{A}}.
\label{initialchoice}
\end{align}
We also define the matrix $\bm{\Psi}$ as follows.
\begin{align}
    \bm{\Psi}  =  & \left( \sum_{l=0}^{\delta_1 - 1} \bm{A}_1 ^l \right)   \bm{B}_{1} 
                + \bm{A}_1^{\delta_1}  \left( \sum_{l=0}^{\delta_2 - 1} \bm{A}_2 ^l \right)  \bm{B}_{2} \nonumber \\
                 & +  \bm{A}_1^{\delta_1} \bm{A}_2^{\delta_2} \left( \sum_{l=0}^{\delta_3 - 1} \bm{A}_3 ^l \right) \bm{B}_{3} \nonumber   \\
               & + \ldots + \bm{A}_1^{\delta_1} \bm{A}_2^{\delta_2} \cdots \bm{A}_{I-1}^{\delta_{I-1}} (\bm{I} - \bm{A}_I)^{-1} \bm{B}_I.  
                \label{Psi} 
\end{align} 
The following theorem provides an expression for obtaining the row vector $\bm{\kappa}$. 
\begin{theorem}
\label{theorem:thm2}
    Consider the $I$-regime AMC $X_k$, $k \geq 0$, constructed for the age-dependent server selection problem, characterized with a set of thresholds $\{ \tau_j \}_{j=1}^J$,
    transient matrices given in \eqref{DefA1} and \eqref{DefAi}, and absorption matrices in \eqref{DefBi}. Then, the probability vector $\bm{\kappa}$ is the stationary solution of a DTMC with probability transition matrix $\bm{B}=\bm{A}\bm{\Psi}$,
\begin{align}
    \bm{\kappa} & = \bm{\kappa}\bm{B}, \ \bm{\kappa}\bm{1} = 1, \label{mainresultthm2}
\end{align}
from which the initial probability vector $\bm{\beta}$ of the MR-AMC $X_k$ is expressed as in \eqref{initialchoice}, which completes the 4-tuple characterization of the MR-AMC $X_k$.
\end{theorem}
The proof of Theorem~\ref{theorem:thm2} is given in Appendix~\ref{appendix:B}.

% Notice that the absorbing state $3$ can only be reached from regime-$3$ when $P_2$ completes on server-$2$. 
% Moreover, the probability of absorption into state 2 is the same as $\kappa_2$.
% Therefore, from \eqref{absorptionfinal}, we have the following identity for $\kappa_2$,
% \begin{align}
%    \kappa_{2} & = \bm{\beta}_3 (\bm{I} - \bm{A}_3)^{-1}  \bm{B}_{3,2}, \label{beta12} \\
%    & = \bm{\beta}  \underbrace{\bm{A}_1^{\tau_1} \bm{A}_2^{\tau_2-\tau_1}(\bm{I} - \bm{A}_3)^{-1}  \bm{B}_{3,2}}_{\bm{c}}. \label{beta12son}
% \end{align}
% Thus, we have the following closed-form expression for $\kappa_2$,
% \begin{align}
%      \kappa_{2} & = \frac{\bm{\alpha}_1 \bm{c}_1}{1 + \bm{\alpha}_1 \bm{c}_1 - \bm{\alpha}_2 \bm{c}_2}, 
%      \label{kappa2}
% \end{align}
% where $\bm{c}_1$ is the column vector corresponding to the first $M_1$ entries of $\bm{c}$, and  $\bm{c}_2$ is composed of its next $M_2$ entries. 
At this stage, we have obtained the 4-tuple \eqref{4tuple} that completely characterizes the MR-AMC $X_k$ given the policy $\mathcal{P}$. One can then obtain its transient vector $\bm{x}_k$ according to \eqref{xk} and \eqref{betaevolution}. The following theorem links the distribution of AoI to the transient vector of the MR-AMC $X_k$. 
% provides the relationship between the pmf of  $\Delta$ to the transient probability vector $\bm{x_k}$ of the MR-AMC $X_k$ in the theorem given below. 
%describes the relationship between the distribution of the steady-state AoI $\Delta$ and the MR-AMC $X_k$.
\begin{theorem}
\label{theorem:thm3}
Consider the MR-AMC $X_k$ characterized with the 4-tuple \eqref{4tuple} according to Theorems~\ref{theorem:thm1} and \ref{theorem:thm2}. Then, the following relationship holds between the pmf of the steady-state AoI, $\Delta$, and the transient probability vector $\bm{x}_k, k=1,2,\ldots$ of the MR-AMC $X_k$:
\begin{align}
    p_{\Delta}(n) \propto \Pr (X_n \in \mathcal{C} ) = \bm{x}_n {\bm h}, 
\end{align}
where $\mathcal{C} = \{ (2,\cdot,\cdot)\}$, and
$\bm{h}$ is a $M \times 1$ column vector whose last $M_1+M_2$ entries are one, and zero otherwise.  
\end{theorem}
The proof of Theorem~\ref{theorem:thm3} is given in Appendix~\ref{appendix:C}.

As an immediate outcome of Theorem~\ref{theorem:thm3}, we write the pmf of the AoI as,
\begin{align}
    p_{\Delta}(n) & = \frac{ \bm{\beta}_i \bm{A}_i^{n-\tau_{i-1}}\bm{h}}{\underbrace{ \bm{\beta}_I (\bm{I} - \bm{A}_I)^{-1} \bm{h} + \sum_{i=1}^{J} \sum_{m=\tau_{i-1}}^{\tau_i} \bm{\beta}_i \bm{A}_i^{m-\tau_{i-1}} \bm{h}}_{\eta^{-1}}} \ ,
\end{align}
when $\tau_{i-1} \leq n < \tau_{i}$, and $\eta$ is the proportionality constant described in Theorem~\ref{theorem:thm3}. Given an arbitrary function $f(\cdot)$ of AoI with finite $\mathbb{E}[f(\Delta)]$, the AoI cost can numerically be obtained through \eqref{AoICost}. In some cases, one can obtain the AoI cost in closed form due to the matrix-geometric form of the AoI pmf in final regime-$I$.
For example, when $f(\Delta)=\Delta$, one can write the AoI cost as the sum of two terms, the former being a finite sum, and the latter being an infinite sum,
\begin{align}
 C_A & = \mathbb{E}[\Delta] =  \sum_{n=1}^{\tau_I-1} n \, p_{\Delta}(n) + \underbrace{\sum_{n=\tau_I}^{\infty} 
 n \, p_{\Delta}(n)}_{\Lambda},
\end{align}
where $\Lambda$ can be written in closed-form as,
\begin{align*}
  \Lambda & = \kappa \bm{\beta }_I \left( \tau_I   + (\tau_I+1) \bm{A}_I +  (\tau_I+2) \bm{A}_I^2 + \ldots \right) \bm{h}, \\ 
  & = \kappa \bm{\beta}_I \left( \tau_I  (\bm{I} - \bm{A}_I)^{-1}
  +  (\bm{I} - \bm{A}_I)^{-2} \bm{A}_I \right)  \bm{h}.
\end{align*}

In order to find the transmission cost $C_T$, we first find the probability $p_{W}$ that the update system waits without an ongoing transmission. Since waiting is incurred when the AoI process satisfies $\Delta_k < \tau_1$, we write, 
\begin{align}
    p_{W} & = \sum_{n=1}^{\tau_1 - 1} p_{\Delta}(n).
\end{align}
Subsequently, frequency of server-$j$ transmissions, namely $f_j$, can be written as,
\begin{align}
    f_j & = (1-p_W) \frac{\kappa_j}{\sum_{l=1}^J\kappa_{l} \mathbb{E}[S_{l}]},
\end{align}
from which one can write the transmission cost $C_T$ according to \eqref{TransmissionCost}.
\section{Numerical Examples}
\label{sec:numerical}
In the numerical examples, we use four different servers whose parameters are given in Table~\ref{tab:servers}. The servers $M_1$ and $M_2$ are representative of slower servers and they have mixed geometric service time distributions $\text{MG}(1/100,1/20,0.5,0.5)$ and $\text{MG}(1/70,1/20,0.5,0.5)$, respectively. The service time of the moderately fast server $G$ is geometrically distributed with parameter $1/30$. The final server $U$ is the fastest of all four, and has a service time uniformly distributed in the interval $[12,18]$. 
Moreover, the variability of the server $U$ is the lowest, and of the server $M_1$ is the highest, expressed in terms of the squared coefficient of variation SCoV defined as the ratio of the variance to the squared mean.  
On the other hand, the cost of using the server $U$ is the highest, and costs of the servers $M_1$ and $M_2$ are the lowest, expressed in terms of the cost parameter $c$. All three servers are of DPH-type and their DPH representations can be obtained as in Subsection~\ref{subsection:dph}. We define the set of all servers as $\mathcal{S}=\{ M_1,M_2,G,U \}$.
We use the notation $S(\tau_1)$ to refer to a single-server system using only server $S \in \mathcal{S}$ and $\tau_1$ is the threshold to be used for wait and transmit decisions.  
When two servers are used, we resort to the notation 
$[S_1,S_2](\tau_1,\tau_2)$ when we wait if the AoI is strictly below $\tau_1$, we transmit over server $S_1$ when the AoI ranges between $\tau_1$ and $\tau_2$, and we transmit over server $S_2$, otherwise, for $S_i \in \mathcal{S}, i=1,2$. 
Similarly, we use the notation $[S_1,S_2,S_3](\tau_1,\tau_2,\tau_3)$ for a three-server system with
$S_i \in \mathcal{S}, i=1,2,3$. 
\begin{table}
\begin{center}
\caption{Parameters of the three servers used in the numerical examples}
\begin{tabular}{|c|c|c|c|c|c|} 
 \hline
 Server & Type & Order & Mean & SCoV & Cost\\ 
 \hline\hline
 $M_1$ & $\text{MG}$ & $2$ & $60$ &  $1.8722$ & $10$ \\ 
 \hline
 $M_2$ & $\text{MG}$ & $2$ & $45$ &  $1.5951$ & $10$ \\ 
 \hline
 $G$ & $\text{Geo}$ & $1$  & $30$ & $.9667$ & $100$ \\
 \hline
 $U$ & $\text{Unif}$ & $18$ & $15$ & $.0178$ & $500,1500$ \\
 \hline
\end{tabular}
\end{center}
\label{tab:servers}
\end{table}
\begin{figure}[ht]
	\centering
	\includegraphics[width=0.9 \columnwidth]{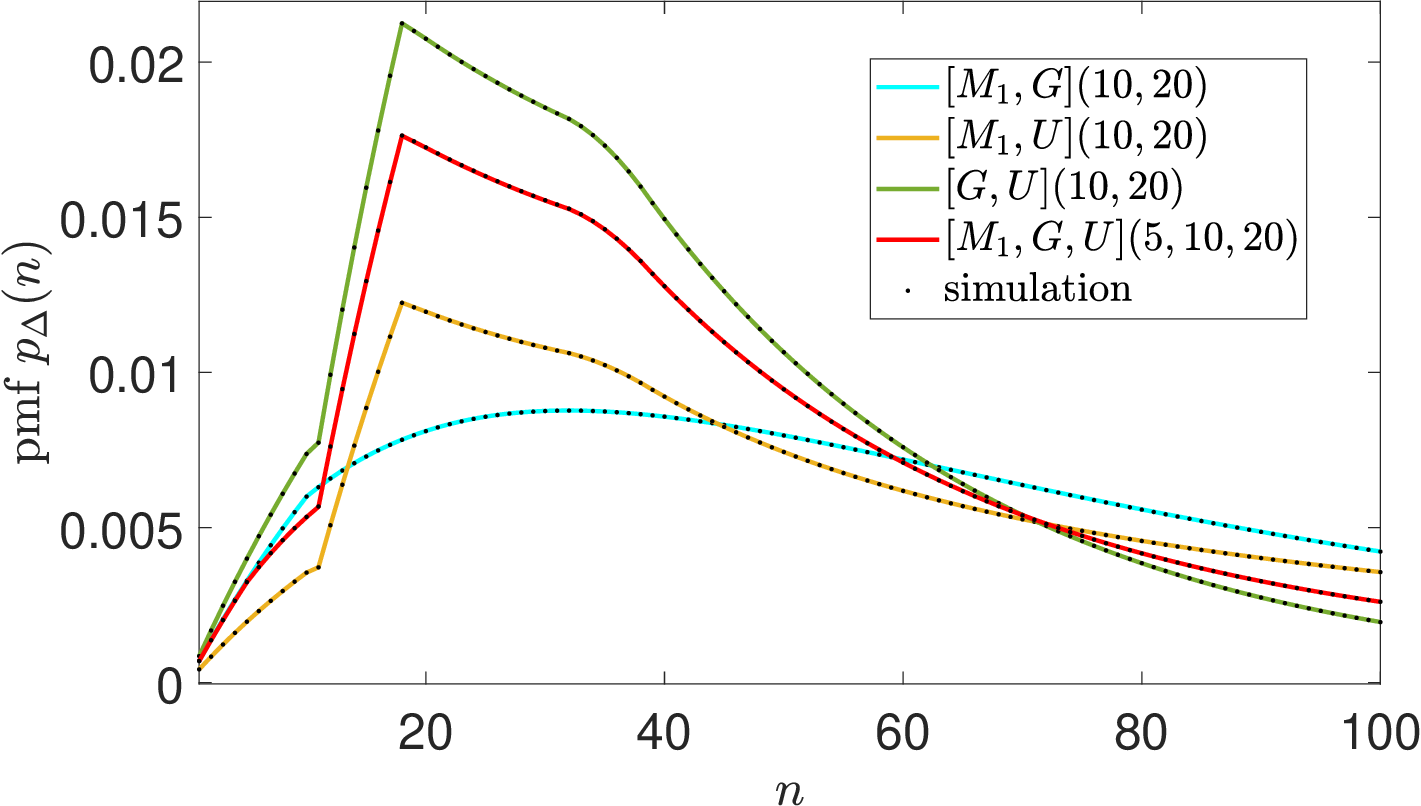}
	\caption{The pmf of the AoI process $p_{\Delta}(n)$ when the three dual-server policies $[M_1,G]$, $[M_1,U]$, and $[G,U]$ are used with the two thresholds $\tau_1=20,\tau_2=50$, along with the triple-server policy $[M_1,G,U]$ policy with $\tau_1=5,\tau_2=10,\tau_3=20$. Simulation results are depicted with black dots.}
	\label{fig:ValidateDistribution}
\end{figure}

\begin{figure}[tbh]
	\centering
	\includegraphics[width=\columnwidth]{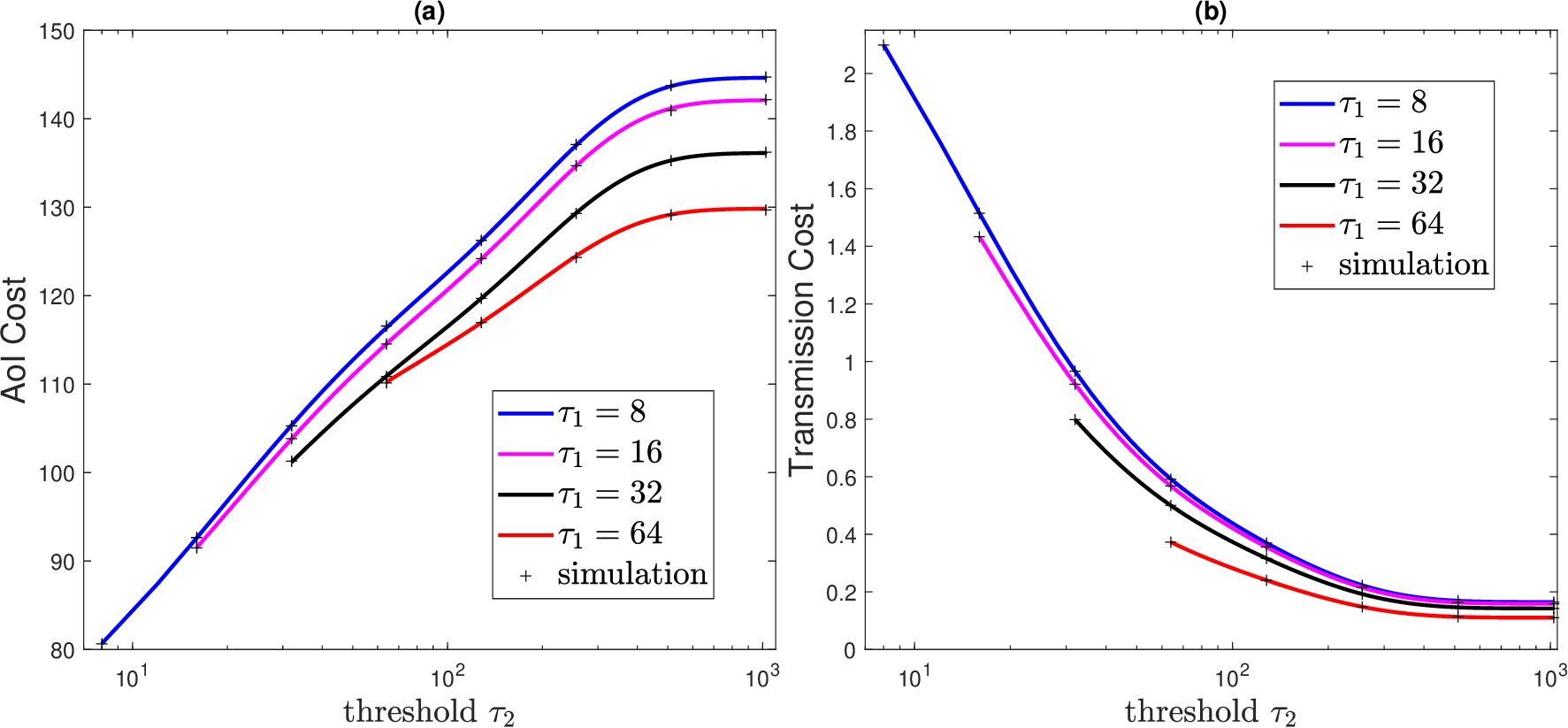}
	\caption{(a) AoI cost (b) transmission cost, depicted as a function of the threshold $\tau_2 \geq \tau_1$, for the dual-server policy $[M_1,G](\tau_1,\tau_2)$ obtained with analysis and simulations (depicted by the marker +) for four different values of $\tau_1$.}
	\label{fig:ValidateTransmissionCost}
\end{figure}

In the first numerical example, we employ the three dual-threshold systems $[M_1,G]$, $[M_1,U]$ and 
$[G,U]$ by fixing the thresholds to $\tau_1=10, \ \tau_2=20$. Also, we study the triple-threshold system $[M_1,G,U]$
when  $\tau_1=5, \ \tau_2=10, \ \tau_3=20$.
Then, we analytically obtain the pmf $p_{\Delta}(n)$ for all the studied policies, and also obtain the AoI pmf by simulations with a simulation time of $5 \times 10^8$ time slots.
The pmf results obtained with the analytical model and simulations given in Fig.~\ref{fig:ValidateDistribution} match perfectly, validating the proposed MR-AMC based approach. As expected, when faster servers are used with lower variability, then the corresponding AoI distribution becomes more concentrated at lower values of age. However, recall that such servers are more costly to use, whose impact on system performance is studied in the next example in which we focus only on the policy $[M_1,G](\tau_1,\tau_2)$ for four different values of $\tau_1=8,16,32,64$ and we take the AoI cost as $C_A = \mathbb{E}[\Delta]$.
We then depict the AoI cost $C_A$ and the transmission cost $C_T$ in two separate sub-figures 
in Fig.~\ref{fig:ValidateTransmissionCost} as a function of the second threshold $\tau_2$.
We observe that the analytical and simulation results perfectly match for both the AoI and transmission costs. As $\tau_2$ increases, the system is less likely to transmit over the faster server $G$ which has a higher cost than server $M_1$, as a result of which the AoI cost $C_A$ and the transmission cost $C_T$ increase and decrease, respectively, with increasing $\tau_2$. Moreover, an increase of the parameter $\tau_1$ increases the likelihood of idle slots, thereby reducing transmission costs. 

\begin{figure}
\centering
	\includegraphics[width=0.75\columnwidth]{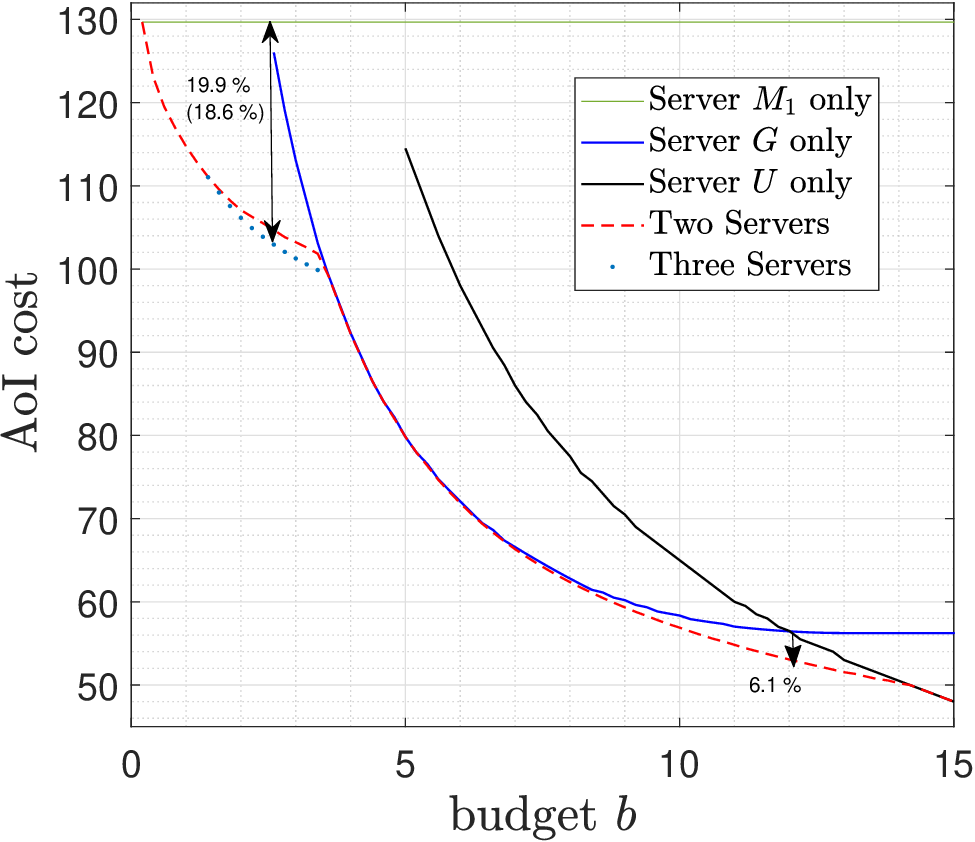}
	\caption{Minimum attainable cost in Scenario 1 for a given transmission cost budget $b$ under various policies}
	\label{fig:minimumAoIa}
\end{figure}

\begin{figure}
\centering
	\includegraphics[width=0.7\columnwidth]{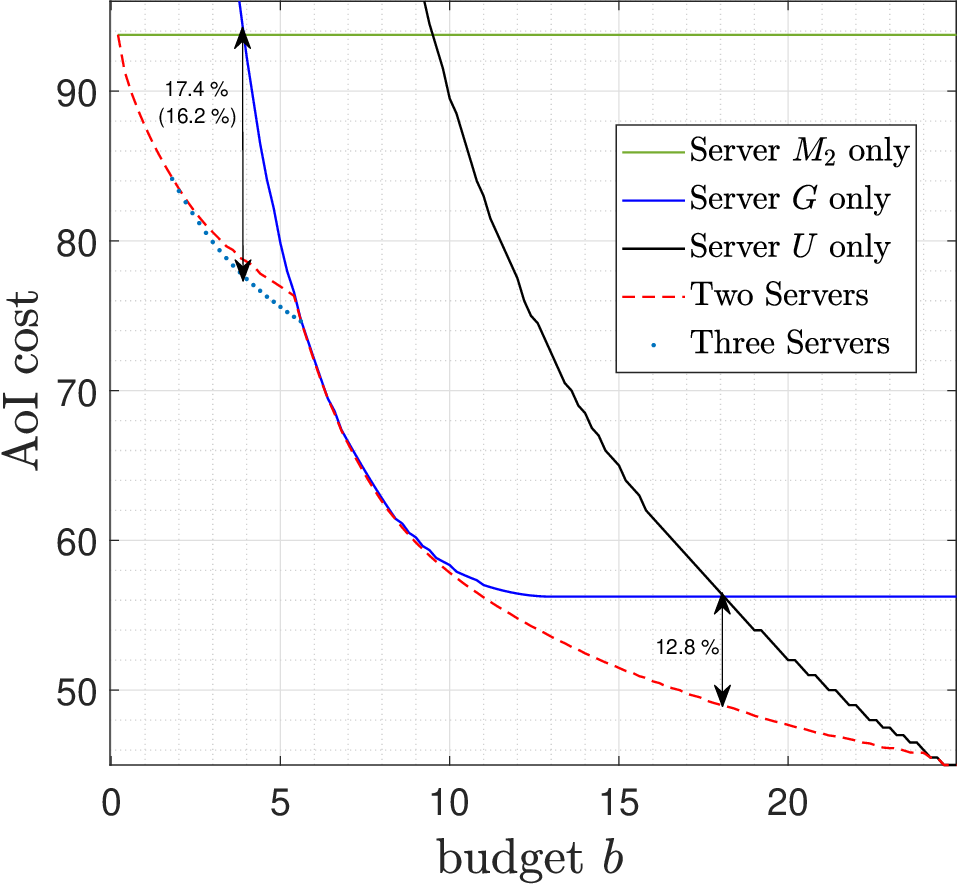}
	\caption{Minimum attainable cost in Scenario 2 for a given transmission cost budget $b$ under various policies}
	\label{fig:minimumAoIb}
\end{figure}

In Scenario 1 of the final numerical example, we  use the set of servers $\{ M_1,G,U\}$ with the transmission cost parameter of $U$, $c_U$, is set to 500 as in the previous examples. Then, for a given transmission cost budget $b$, we study a given policy for all possible thresholds up to $\tau_{max}=200$, and find the thresholds resulting in the minimum AoI cost (taken as average AoI) among the ones whose transmission costs do not exceed $b$. For a single-server only policy, we only find the threshold $\tau_1$. For the {\em Two Servers} policy, we are allowed to use {\em at most} two servers while performing age-dependent server selection. Particularly, we analyze the three dual-server policies $[M_1,G]$, $[M_1,U]$, and $[G,U]$ for all possible pairs of thresholds $(\tau_1,\tau_2)$ along with the three single-server policies, and choose one of the above six policies resulting in the minimum AoI, for a given budget $b$.
On the other hand, in the {\em Three Servers} policy, one can use up to three servers {\em at most}.
For this purpose, we analyze the triple-server policy
$[M_1,G,U]$ for all possible threshold 3-tuples $(\tau_1,\tau_2,\tau_3)$, and choose the particular value of thresholds resulting in the minimum AoI, for a given budget $b$. We choose this policy if it gives lesser AoI than the Two Servers policy for a given budget $b$.  
Our results are depicted in Fig.~\ref{fig:minimumAoIa} in which the AoI cost is plotted as a function of the budget parameter $b$ for various policies whereas for the Three Servers policy, results are only depicted when they resulted in lesser AoI cost than the Two Servers policy. We observe that for lower values of the budget $b$, substantial reductions up to $18.6 \%$ in average AoI are possible by using age-dependent server selection using the policy $[M_1,G]$ against the best single-server policy.
In the same regime, the AoI cost can further be reduced by using age-dependent server selection with the Three Servers policy $[M_1,G,U]$. In particular, the largest reduction in average AoI with with the Three Servers policy 
is found as $19.9 \%$. For larger values of $b$, we observed AoI cost reductions up to $6.1 \%$ with the $[G,U]$ policy but we did not observe benefits of using the Three Servers policy, in this regime. 
In Fig.~\ref{fig:minimumAoIb}, we repeat the same experiment for a separate case, called Scenario 2, by replacing the server $M_1$ by $M_2$ and also the parameter $c_U$ is changed to 1500. While doing so, the gap between the first and second servers is reduced in terms of server rate, whereas the gap between the second and third servers, is increased in terms of the transmission cost. 
We have similar observations with Scenario 1, but the AoI cost reduction is smaller in the first regime whereas in the second regime, it is more significant. As a general observation, employing age-dependent server selection results in more significant average AoI reduction when the diversity among the servers is more pronounced in terms of service rates and costs. Moreover, we are inclined to believe that most of the gains in AoI performance stem from the use of age-dependent server selection among two servers, and gains with the use of three servers appear to be rather limited.

\section{Conclusions}
\label{sec:conclusions}
In this paper, we proposed a novel method to obtain the distribution of AoI in a single-source multi-server generate-at-will discrete-time status update system with DPH-distributed service times, where a multi-threshold policy is imposed to determine whether to wait or transmit, and if latter, through which server to transmit.
For this purpose, multi-regime absorbing Markov chains are employed for modeling the distribution of AoI, which have not been explored in the literature, to the best of our knowledge.
This exact analytical model enables one to find the optimum thresholds under which the average of an arbitrary function of AoI is minimized under a constraint on the overall transmission costs.
The model is validated with simulations for dual- and triple-server scenarios for which the benefits of using optimum multi-threshold policies are demonstrated. 
We have shown that up to $19.9 \%$ reduction in average AoI is possible with the proposed framework, as opposed to using one server only. Moreover, the majority of the performance gain stems from age-dependent server selection among two servers, and the contribution of using three servers appears to be relatively limited. 
Modeling of random arrival scenarios, continuous-time settings, and service time distributions of non-renewal type, can be considered for future work.
\appendices
\section{Proof of Theorem~\ref{theorem:thm1}}
\label{appendix:A}
We first define the transient probability vector $\bm{x}_k$ and absorption probability vector $\bm{\tilde{x}}_k$ at time $k$ as,
\begin{align}
   \bm{x}_k & = \begin{pmatrix}
      x_{k,1}  & \cdots & x_{k,M}  
   \end{pmatrix}, \;
   x_{k,m}   = \Pr (X_k=m), \label{Defxk} \\
    \bm{\tilde{x}}_k & = \begin{pmatrix}
      \tilde{x}_{k,1} & \cdots & \tilde{x}_{k,J}  
   \end{pmatrix}, \;
   \tilde{x}_{k,j}   = \Pr (X_k=M+j). \label{yk}
\end{align} 
Let us first focus on the transitions in regime-$1$ for which case we have,
\begin{align}
  \begin{pmatrix}
      \bm{x}_k & \bm{\tilde{x}}_k 
  \end{pmatrix}  & = \begin{pmatrix}
      \bm{\beta}_1 & \bm{0}
  \end{pmatrix} \left( \begin{array}{c|c} \bm{A}_1 & \bm{B}_1 \\ \hline \bm{0} & \bm{I} \end{array} \right)^k, \,  0 \leq k \leq \tau_1.
\end{align}
Therefore, for $0 \leq k \leq \tau_{1}$,
\begin{align}
\bm{x}_k & = \bm{\beta}_1 \bm{A}_1^{k}, \label{xkR1}  
\quad \bm{\tilde{x}}_k  = \bm{\beta}_1 \left( \sum_{l=0}^{k-1}  \bm{A}_1^{l} \right) \bm{B}_1. 
% \\ 
%          & = \bm{\beta}_1   (\bm{I} - \bm{A}_1^k) (\bm{I} - \bm{A}_1)^{-1} \bm{B}_1.
\end{align}
Since $\tilde{x}_{\tau_1,j}$ is the probability that absorption occurs into absorbing state-$j$ from a transition in regime-$1$, we have
\begin{align}
    \bm{\sigma}_1 & = \bm{\tilde{x}}_{\tau_1} = 
                   \bm{\beta}_1 \left( \sum_{l=0}^{\delta_1-1} \bm{A}_1^l \right)  \bm{B}_1. \label{gammaR1}
              %    & = \bm{\beta}_1  (\bm{I} - \bm{A}_1^{\delta_1})  (\bm{I} - \bm{A}_1)^{-1}           
\end{align}
Let us now consider regime-$2$. In this regime, for $\tau_1 \leq k \leq \tau_2$,
\begin{align}
  \begin{pmatrix}
      \bm{x}_k & \bm{\tilde{x}}_k 
  \end{pmatrix}  & = \underbrace{\begin{pmatrix}
      \bm{x}_{\tau_1} & \bm{\tilde{x}}_{\tau_1}
  \end{pmatrix}}_{\begin{pmatrix}
      \bm{\beta}_2 & \bm{\sigma}_1
  \end{pmatrix}}  \left( \begin{array}{c|c} \bm{A}_2 & \bm{B}_2 \\ \hline \bm{0} & \bm{I} \end{array} \right)^{k-\tau_1}, 
    % & =  \left( \begin{array}{c|c} \bm{A}_2 & \bm{B}_2 \\ \hline \bm{0} & \bm{I} \end{array} \right)^{k-\tau_1}, \, \tau_1 \leq k \leq \tau_2,
\end{align}
which can be shown by \eqref{betaevolution} and \eqref{gammaR1}.
Consequently, for $\tau_1 \leq k \leq \tau_{2}$, we have,
\begin{align}
\bm{x}_k & = \bm{\beta}_2 \bm{A}_2^{k-\tau_1}, \label{xkR2} \\  
\bm{\tilde{x}}_{k} & = \bm{\sigma}_1 + \bm{\beta}_2 \left( \sum_{l=0}^{k-\tau_1-1}  \bm{A}_2^{l} \right) \bm{B}_2.  
% \\ 
         % & = \bm{\sigma}_1 + \bm{\beta}_1   (\bm{I} - \bm{A}_2^{k-\tau_1}) (\bm{I} - \bm{A}_2)^{-1} \bm{B}_2.
    \label{ykR2} 
\end{align}
Since $\tilde{x}_{\tau_2,j}$ is the probability that absorption occurs into absorbing state-$j$ from a transition in regime-$1$ or regime-$2$, we have,
\begin{align}
    \bm{\sigma}_2 & = \bm{\tilde{x}}_{\tau_2} - \bm{\sigma_1}, \\
                  & = \bm{\beta}_2   \left( \sum_{l=0}^{\delta_2-1}  \bm{A}_2^{l} \right) \bm{B}_2.
                  % (\bm{I} - \bm{A}_2^{\delta_2}) (\bm{I} - \bm{A}_2)^{-1} \bm{B}_2.
                  \label{gammaR2}
\end{align}
If we continue the same analysis for the other regimes, we obtain, 
\begin{align}
\bm{x}_k & = \bm{\beta}_i \bm{A}_i^{k-\tau_{i-1}}, \ \tau_{i-1} \leq k < \tau_{i}, \label{xk}
\end{align}
which is shown to hold for the first two regimes in \eqref{xkR1} and \eqref{xkR2}.
With the same analysis for all the regimes, we obtain  the expression for the absorption probability vector $\bm{\sigma}_i$ for regime-$i$ in \eqref{absorptionall} which are also obtained for the first two regimes in \eqref{gammaR1} and \eqref{gammaR2}. It is not difficult to obtain \eqref{absorptionfinalregime} from \eqref{absorptionall} via the observation $\lim_{l \rightarrow \infty}\bm{A}_I^l = \bm{0}$ since $\bm{A}_I$ is a sub-stochastic matrix with all its eigenvalues being strictly inside the unit circle. Finally, 
\begin{align}
    F_T(n) & = 1 - \bm{x}_n \bm{1}, \\
        & = 1 - \bm{\beta}_i \bm{A}_i^{n-\tau_{i-1}} \bm{1}, \, \tau_{i-1} \leq n < \tau_{i},
\end{align}
from \eqref{xk}, which completes the proof.
\section{Proof of Theorem~\ref{theorem:thm2}}
\label{appendix:B}
Considering the MR-AMC $X_k$, the probability that $P_2$ is served by server-$j$, also the probability of absorption 
into absorption state-$j$, is also equal to the probability that $P_1$ is served by server-$j$, $\kappa_j$, since $P_1$ and $P_2$ are any two successive transmitted packets.
Recalling the definition of per-regime absorption probability vector $\bm{\sigma}_i$ for the MR-AMC $X_k$ from Theorem~\ref{theorem:thm1}, the row vector $\bm{\kappa}$ satisfies the following,
\begin{align}
    \bm{\kappa} & = \sum_{i=1}^I \bm{\sigma}_i =  \bm{\beta} \bm{\Psi} = \bm{\kappa} \bm{B}.
\end{align} 
Due to the definition of $\bm{Q}_i$, $\bm{B}_i \bm{1} = \bm{1} -\bm{A}_i \bm{1}$. 
Therefore,
\begin{align*}
    \left( \sum_{l=0}^{\delta_i - 1} \bm{A}_i^l \right) \bm{B}_i \bm{1} & = (\bm{I} - \bm{A}_i^{\delta_i}) \bm{1}. 
\end{align*}
% Also, the matrix 
% $(\bm{I} - \bm{A}_i)^{-1}$, also known as the fundamental matrix of the associated AMC (see \cite{kemeny1960finite}), is a non-negative matrix.  
% Moreover,
% $(\bm{I} - \bm{A}_i)^{-1} \bm{B}_{i} \bm{1} = \bm{1}$ which leads us to the following expression for the matrix product 
% $\bm{\Psi} \bm{1}$, 
Consequently,
\begin{align*}
     \bm{\Psi} \bm{1} = & (\bm{I}-  \bm{A}_1^{\delta_1})  \bm{1}  
               + \bm{A}_1^{\delta_1}  (\bm{I}-  \bm{A}_2^{\delta_2}) \bm{1}  
               +  \bm{A}_1^{\delta_1} \bm{A}_2^{\delta_2} (\bm{I}-  \bm{A}_3^{\delta_3})  \bm{1} \\
              & + \ldots +  \bm{A}_1^{\delta_1} \bm{A}_2^{\delta_2} \cdots \bm{A}_{I-1}^{\delta_{I-1}} \bm{1} = \bm{1}.
                \label{Psi1} 
\end{align*} 
It is also clear that $\bm{A}\bm{1}=\bm{1}$ since $\bm{\alpha}_i \bm{1}=1$, which proves that the row sums of the matrix $\bm{B}$ are one, i.e., $\bm{B}\bm{1}=\bm{A}\bm{\Psi}\bm{1} = \bm{1}$. Also note that in the construction of the matrix $\bm{B}$, we always add and multiply non-negative numbers. Therefore, the matrix $\bm{B}$ is a probability transition matrix whose stationary solution is given by \eqref{mainresultthm2}, which completes the proof. 
%                 (\bm{I} - \bm{A}_1)^{-1} (\bm{I}-  \bm{A}_1^{\delta_1}) \bm{B}_{1} \nonumber \\
%                & + \bm{\beta}  \bm{A}_1^{\delta_1} (\bm{I} - \bm{A}_2)^{-1} (\bm{I}-  \bm{A}_2^{\delta_2}) \bm{B}_{2} \nonumber \\
%                 & + \bm{\beta}   \bm{A}_1^{\delta_1} \bm{A}_2^{\delta_2}(\bm{I} - \bm{A}_3)^{-1} (\bm{I}-  \bm{A}_3^{\delta_3})  \bm{B}_{3} \nonumber \\
%                 & \, \, \, \vdots \nonumber  \\
%                & + \bm{\beta}  \bm{A}_1^{\delta_1} \bm{A}_2^{\delta_2} \cdots \bm{A}_{I-1}^{\delta_{I-1}} (\bm{I} - \bm{A}_I)^{-1} \bm{B}_I  \nonumber \\
%                 & =:  \bm{\kappa} \bm{A} \bm{\Psi}
%                 \label{absorptionall} 
% \end{align} 
\section{Proof of Theorem~\ref{theorem:thm3}}
\label{appendix:C}
Revisiting Fig.~\ref{fig:samplepath}, a given AoI cycle-$\ell$ starts with the reception of a packet $P_1$ and continues until the reception of the next packet $P_2$. On the other hand, the AMC $X_k$ starts operation at $k=0$ with the start of service of packet $P_1$, and continues until the reception of the packet $P_2$, at which point it is absorbed into one of the $J$ absorbing states.  After $P_1$ is received, there are two possibilities; the AMC $X_k$ is either visiting the state $(2,l,m), l=1,\ldots,L$, $m=1,\ldots,M_l$, denoted by $\mathcal{C}$, or is absorbed. 
Using a sample path argument, for each AoI cycle, we have a corresponding MR-AMC cycle, and we observe that the AoI cycles in Fig.~\ref{fig:samplepath} and parts of the corresponding MR-AMC cycles that are spent in states in $\mathcal{C}$ overlap. 
Therefore, if the AoI value $n$ is visited in an AoI cycle, then at time $n$, a state in $\mathcal{C}$ is to be visited in the corresponding MR-AMC cycle. 
As a result, the pmf $p_{\Delta}(n)$ turns out to be the same as the probability that $X_n \in \mathcal{C}$ divided by the mean AoI cycle length, which completes the proof.


\bibliographystyle{IEEEtran}
\bibliography{bibl}

@inproceedings{atasayar_etal_mobihoc25,
author = {Adem Utku Atasayar and Aimin Li and \c{C}a\u{g}r\i{} Ar\i{} and Elif Uysal},
title = {Fresh Data Delivery: Joint Sampling and Routing for Minimizing the Age of Information},
year = {2025},
month ={October},
address = {Houston, TX, USA},
booktitle = {Proceedings of MobiHoc},
pages = {291–300},
}

@ARTICLE{dogan_akar_tcom21,
  author={Doǧan, Ozancan and Akar, Nail},
  journal={IEEE Transactions on Communications}, 
  title={The Multi-Source Probabilistically Preemptive {M/PH/1/1} Queue With Packet Errors}, 
  year={2021},
  volume={69},
  number={11},
  pages={7297-7308}
  }

@INPROCEEDINGS{banerjee_ulukus_isit24,
  author={Banerjee, Subhankar and Ulukus, Sennur},
  booktitle={IEEE International Symposium on Information Theory (ISIT)}, 
  title={When to Preempt in a Status Update System?}, 
  year={2024},
  volume={},
  number={},
  pages={1379-1384}
  }

@ARTICLE{cosandal_etal_tit25,
  author={Cosandal, Ismail and Akar, Nail and Ulukus, Sennur},
  journal={IEEE Transactions on Information Theory}, 
  title={Multi-Threshold {AoII}-Optimum Sampling Policies for Continuous-Time {Markov} Chain Information Sources}, 
  year={2025},
  volume={71},
  number={9},
  pages={6968-6988}
  }

@article{albrecher2019inhomogeneous,
  title={Inhomogeneous phase-type distributions and heavy tails},
  author={Albrecher, Hansj{\"o}rg and Bladt, Mogens},
  journal={Journal of Applied Probability},
  volume={56},
  number={4},
  pages={1044--1064},
  year={2019},
  publisher={Cambridge University Press}
}

@article{albrecher2022fitting,
  title={Fitting inhomogeneous phase-type distributions to data: the univariate and the multivariate case},
  author={Albrecher, Hansj{\"o}rg and Bladt, Mogens and Yslas, Jorge},
  journal={Scandinavian Journal of Statistics},
  volume={49},
  number={1},
  pages={44--77},
  year={2022},
  publisher={Wiley Online Library}
}

@article{yates2019,
  author={Yates, R. D. and Kaul, S. K.},
  journal={IEEE Trans. Inf. Theory}, 
  title={The Age of Information: Real-Time Status Updating by Multiple Sources}, 
  month={March},
  year={2019},
  volume={65},
  number={3},
  pages={1807-1827}}

@INPROCEEDINGS{purdue_paper,
  author={Lee, Won Jun and Wang, Chih–Chun},
  booktitle={IEEE International Symposium on Information Theory (ISIT)}, 
  title={{AoI}-optimal Scheduling for Arbitrary {K-channel} Update-Through-Queue Systems}, 
  year={2024},
  volume={},
  number={},
  pages={957-962},
  doi={10.1109/ISIT57864.2024.10619623}}

@ARTICLE{sun_etal_TIT17,
  author={Sun, Yin and Uysal-Biyikoglu, Elif and Yates, Roy D. and Koksal, C. Emre and Shroff, Ness B.},
  journal={IEEE Transactions on Information Theory}, 
  title={Update or Wait: How to Keep Your Data Fresh}, 
  year={2017},
  volume={63},
  number={11},
  pages={7492-7508}
  }

@article{akar_gamgam_comlet23,
  title={Distribution of Age of Information in Status Update Systems with Heterogeneous Information Sources: An Absorbing {M}arkov Chain-based Approach},
  author={Akar, N. and Gamgam, E. O.},
  journal={IEEE Commun. Lett.}, 
  volume={27},
  number={8},
  pages={2024-2028},
  month={May},
  year={2023}}

@article{bagui2024stirling,
  title={The {S}tirling numbers of the second kind and their applications},
  author={Bagui, S. and Mehra, K.L.},
  journal={Ala. j. math.},
  volume={47},
  number={1},
  pages={1--22},
  year={2024}
}

@book{nielsen_book,
	author               = {Mogens Bladt and Bo Friis Nielsen},
	publisher            = {Springer},
	title                = {Matrix-Exponential Distributions in Applied Probability},
	year                 = {2017},
        address = {New York, NY, USA}
}

@ARTICLE{pan_etal_tmc22,
  author={Pan, Jiayu and Bedewy, Ahmed M. and Sun, Yin and Shroff, Ness B.},
  journal={IEEE Transactions on Mobile Computing}, 
  title={Age-Optimal Scheduling Over Hybrid Channels}, 
  year={2023},
  volume={22},
  number={12},
  pages={7027-7043},
  doi={10.1109/TMC.2022.3205292}}

@article{hawking1999methods,
  title={Methods for information server selection},
  author={Hawking, David and Thistlewaite, Paul},
  journal={ACM Transactions on Information Systems (TOIS)},
  volume={17},
  number={1},
  pages={40--76},
  year={1999},
  publisher={ACM New York, NY, USA}
}

@inproceedings{akar_Asilomar,
  title={Age-Dependent Server Selection in a Dual-Server Status Update System},
  author={Akar, N. and Cosandal, I. and Ulukus, S.},
  booktitle={The Asilomar Conference on Signals, Systems, and Computers},
  month={October},
  address = {Monterey, CA, USA},
  year={2025}}

@INPROCEEDINGS{fei_etal_infococom98,
  author={Fei, Z.-M. and Bhattacharjee, S. and Zegura, E.W. and Ammar, M.H.},
  booktitle={Proceedings of IEEE INFOCOM }, 
  title={A novel server selection technique for improving the response time of a replicated service}, 
  year={1998},
  volume={2},
  number={},
  pages={783-791 vol.2},
  doi={10.1109/INFCOM.1998.665101}}

@book{telek_book,
	author               = {László Lakatos and László Szeidl and Miklós Telek },
	publisher            = {Springer},
	title                = {Introduction to Queueing Systems with Telecommunication Applications},
	year                 = {2019},
        edition = {Second},
        address = {New York, NY, USA}
}

@book{alfa_book,
	author               = {Attahiru S. Alfa},
	publisher            = {Springer},
	title                = {Applied Discrete-Time Queues},
	year                 = {2016},
        address = {New York, NY, USA}
}

@INPROCEEDINGS{kaul_etal_infocom12, 
	author={S. {Kaul} and R. {Yates} and M. {Gruteser}}, 
	booktitle={IEEE Infocom}, 
	title={Real-time status: How often should one update?}, 
	year={2012}, 
	month={March}
}

@article{kosta_etal_jsac21,
  title={The age of information in a discrete time queue: Stationary distribution and non-linear age mean analysis},
  author={Kosta, Antzela and Pappas, Nikolaos and Ephremides, Anthony and Angelakis, Vangelis},
  journal={IEEE Journal on Selected Areas in Communications},
  volume={39},
  number={5},
  pages={1352--1364},
  year={2021},
  publisher={IEEE}
}

@article{akar_dogan_iot21,
  title={Discrete-time queueing model of age of information with multiple information sources},
  author={Akar, Nail and Dogan, Ozancan},
  journal={IEEE Internet of Things Journal},
  volume={8},
  number={19},
  pages={14531--14542},
  year={2021},
  publisher={IEEE}
}

@ARTICLE{zhang_etal_tcom25,
  author={Zhang, Tianci and Chen, Zhengchuan and Tian, Zhong and Wang, Min and Zhen, Li and Wu, Dapeng Oliver and Li, Yonghui and Quek, Tony Q. S.},
  journal={IEEE Transactions on Communications}, 
  title={Age of Information in {Internet} of Vehicles: A Discrete-Time Multisource Queueing Model}, 
  year={2025},
  volume={73},
  number={5},
  pages={3298-3317}
}

@ARTICLE{tripathi_modiano_TNET,
  author={Tripathi, Vishrant and Modiano, Eytan},
  journal={IEEE/ACM Transactions on Networking}, 
  title={A {Whittle} Index Approach to Minimizing Functions of Age of Information}, 
  year={2024},
  volume={32},
  number={6},
  pages={5144-5158}
}

@ARTICLE{kam_etal_TIT16,
  author={Kam, Clement and Kompella, Sastry and Nguyen, Gam D. and Ephremides, Anthony},
  journal={IEEE Transactions on Information Theory}, 
  title={Effect of Message Transmission Path Diversity on Status Age}, 
  year={2016},
  volume={62},
  number={3},
  pages={1360-1374}
  }

@ARTICLE{akar_ulukus_tcom25,
  author={Akar, Nail and Ulukus, Sennur},
  journal={IEEE Transactions on Communications}, 
  title={Age of Information in a Single-Source Generate-at-Will Dual-Server Status Update System}, 
  year={2025},
  volume={73},
  number={9},
  pages={7431-7444}}

@article{yates_kaul_TIT19,
  author={Yates, R. D. and Kaul, S. K.},
  journal={IEEE Transactions on Information Theory}, 
  title={The Age of Information: Real-Time Status Updating by Multiple Sources}, 
  month={March},
  year={2019},
  volume={65},
  number={3},
  pages={1807-1827}}

@article{kosta_etal_survey,
	year = {2017},
	volume = {12},
	journal = {Foundations and Trends in Networking},
	title = {Age of Information: A New Concept, Metric, and Tool},
	doi = {10.1561/1300000060},
	issn = {1554-057X},
	number = {3},
	pages = {162-259},
	author = {Antzela Kosta and Nikolaos Pappas and Vangelis Angelakis}}

@INPROCEEDINGS{yates_isit18,
  author={Yates, Roy D.},
  booktitle={IEEE International Symposium on Information Theory (ISIT)}, 
  title={Status Updates through Networks of Parallel Servers}, 
  year={2018},
address = {Vail, CO, USA},
  pages={2281-2285}
}

@ARTICLE{elmagid_commag19,
  author={Abd-Elmagid, Mohamed A. and Pappas, Nikolaos and Dhillon, Harpreet S.},
  journal={IEEE Communications Magazine}, 
  title={On the Role of Age of Information in the {Internet of Things}}, 
  year={2019},
  volume={57},
  number={12},
  pages={72-77}}

@article{abbas_survey23,
title = {A comprehensive survey on age of information in massive {IoT} networks},
journal = {Computer Communications},
volume = {197},
pages = {199-213},
year = {2023},
author = {Qamar Abbas and Syed Ali Hassan and Hassaan Khaliq Qureshi and Kapal Dev and Haejoon Jung}
}

@article{yates_survey,
  title={Age of Information: An Introduction and Survey},
  author={R. D. Yates and Y. Sun and D. R. Brown and S. K. Kaul and E. Modiano and S. Ulukus},
  journal={IEEE Jour. Sel. Areas in Comm.},
  volume={39},
  number={5},
  pages={1183-1210},
  month={May},
  year={2021}}

@article{yates2020age,
  title={The age of information in networks: Moments, distributions, and sampling},
  author={Yates, R. D.},
  journal={IEEE Trans. Inf. Theory}, 
  volume={66},
  number={9},
  pages={5712--5728},
  month={May},
  year={2020}
}

@unpublished{yifan_etal_unpublished25,
title={Absorbing {Markov} Chain-Based Analysis of Age of Information in Discrete-Time Dual-Queue Systems}, 
      author={Yifan Feng and Nail Akar and Zhengchuan Chen and Mehul Motani},
      year={2025},
      note={ArXiv 2509.23360},
      archivePrefix={arXiv}
}
\end{document}